\documentclass[aps,prb,twocolumn,showpacs]{revtex4-1}

\usepackage{graphicx}
\DeclareGraphicsExtensions{.png}
\DeclareGraphicsExtensions{.jpeg}

\usepackage{xcolor}
\usepackage{amsmath}
\usepackage{amssymb}

\usepackage{dcolumn}
\usepackage{bm}
\usepackage{hyperref}

\usepackage[latin1]{inputenc}
\usepackage[american]{babel}
\usepackage{latexsym}
\usepackage{float}

\begin{document}


\title{Electrical Detection of Individual Skyrmions in Graphene Devices}

\index{F. Finocchiaro}
\index{J. L. Lado}
\index{J. Fernandez-Rossier}

\author{F. Finocchiaro$^{1}$, J. L. Lado$^{2}$ and J. Fernandez-Rossier$^{2,3}$}
\affiliation{$^1$IMDEA Nanociencia, Calle de Faraday 9, Cantoblanco 28049 - Madrid, Spain}
\affiliation{$^2$QuantaLab, International Iberian Nanotechnology Laboratory (INL), Av. Mestre Jose Veiga, 4715-310 Braga, Portugal}
\affiliation{$^3$Departamento de Fisica Aplicada, Universidad de Alicante, San Vicente del Raspeig, 03690 Spain}

\date{\today}

\begin{abstract}
We study a graphene Hall probe located on top of
a magnetic surface as a detector of
skyrmions, using as
working principle the anomalous Hall effect produced by the
exchange interaction of the graphene electrons with the non-coplanar
magnetization of the skyrmion. We study the magnitude of the effect as a
function of the exchange interaction, skyrmion size and device dimensions. Our
calculations for multiterminal graphene nanodevices, working in the ballistic
regime, indicate that for realistic exchange interactions a single skyrmion
would give Hall voltages well within reach of the experimental state of the art. The proposed
device could act as an electrical transducer that marks the presence of a
single skyrmion in a nanoscale region, paving the way towards the integration
of skyrmion-based spintronics and graphene electronics. 
\end{abstract}

\maketitle


\section{Introduction}
Skyrmions are magnetic non-coplanar spin textures that are attracting a great
deal of attention for both their appealing physical properties\cite{Roszler2006}
and their potential use in spintronics\cite{Duine13,Rosch2017,Dupe16,Editorial2013}. They have been observed forming
lattices in a variety of non-centrosymmetric magnetic crystals\cite{Muhlbauer09,Munzer10,yu2012skyrmion,tokunaga2015new},
including insulating materials such as the chiral-lattice magnet
Cu$_2$OSeO$_3$\cite{Seki12,Langner14,zhang2016imaging}. They also form two dimensional arrays in
atomically thin layers of Fe deposited on Ir(111)\cite{Heinze11,Romming15}.
In these systems the spins typically feel a competition between aligning
with their neighbors and being perpendicular to them, what favors chiral
ordering. A variety of interactions can assist non-collinear arrangements, 
including Dzyaloshinskii-Moryia interactions, dipolar interactions and frustrated exchange interactions and the size of an individual skyrmion can range from 1 nm to 1 $\mu$m depending on
which specific mechanism is involved. To date, these magnetic structures are
detected by means of neutron scattering\cite{Muhlbauer09}, electron
microscopy\cite{Yu10} and even individually, with atomic scale resolution, by
means of spin polarized scanning tunneling
microscopy\cite{Pfleiderer11,Heinze11} and atomic
size sensors\cite{dovzhenko2016imaging}.

The particle-like nature of skyrmions has motivated proposals to use them as elementary units to store classical digital information, inspired by the magnetic domain-wall racetrack memories\cite{Parkin08}. Such a perspective has become increasingly attractive since it has been experimentally proved\cite{Romming15} the possibility of manipulating two-dimensional magnetic lattices by creating and destroying individual skyrmions by means of spin-polarized currents in STM devices.
This, along with the experimental finding\cite{Jonietz10} of skyrmion motion driven by ultralow current densities of the order of $10^{-6}$ A m$^{-2}$, considerably smaller than those needed for domain wall motion in ferromagnets, makes skyrmions potentially optimal candidates for the next generation of magnetoelectronic readout devices.

Mathematically, skyrmions are topologically non-trivial objects whose topology content is embedded in an index, the winding number $N$, defined as
\begin{equation}
N=\frac{1}{4\pi}\int_{A}\mathbf{n}(x,y)\cdot\left(\frac{\partial\mathbf{n}(x,y)}{\partial x}\times\frac{\partial\mathbf{n}(x,y)}{\partial y}\right)dx\, dy
\label{eq: topo charge}
\end{equation}
where $\mathbf{n}(x,y):\mathbb{R}^2 \rightarrow \mathbb{R}^3$ is a classical magnetization field and the two-dimensional integral is performed over the overall area occupied by the skyrmion. 
The winding number $N$ can only acquire integer values, and a skyrmion is distinguished from other topologically trivial magnetic
 textures for exhibiting a non-zero value of the integer $N$. The magnetization field $\textbf{n}(x,y)$ of a skyrmion can be expressed as a mapping from the polar plane coordinates $\mathbf{r}=\left(r,\phi\right)$ to the unit sphere coordinates $\left(\Phi,\Theta\right)$
\begin{equation}
\mathbf{n}(\mathbf{r})=\left(\cos\Phi(\phi)\sin\Theta(r),\sin\Phi(\phi)\sin\Theta(r),\cos\Theta(r)\right)
\label{eq: n}
\end{equation}
provided the spin configuration at $r=\infty$ is $\phi$-independent so that it can be mapped to a single point on the sphere. The mapping is specified by the two functions\cite{Nagaosa2013}:
\begin{equation}
\Phi(\phi)=N\phi+\gamma
\end{equation}
and $\Theta(r)$ varies from $0$ for large $r$ to $\pi$ as we approach $r=0$, the core of the skyrmion. Here we adopt the following model:
\begin{equation}
\Theta(r)=\left\{ \begin{array}{ll}
\pi & \mbox{for }r=0\\
f(r) = \pi\left(1-r/R\right) & \mbox{for \ensuremath{0<r\leq R}}\\
0 & \mbox{for }r>R
\end{array}\right.
\end{equation}
where $N$ is the skyrmion winding number introduced in (\ref{eq: topo charge}), $\gamma$ is a phase termed helicity that can be gauged away by rotation around the $z$-axis, and $f(r)=\pi\left(1-r/R\right)$ is a function of the radial coordinate that describes a smooth radial profile inside of the skyrmion radius $R$. Such a texture describes a magnetic configuration where the spins are all aligned perpendicular to the film plane with the exception of those comprised within the radius $R$ where they all progressively align along the anti-parallel direction, that is picked up exactly at $r=0$. The condition that the spins at $r=0$ and $r=\infty$ are oppositely oriented is crucial in order to ensure a non-trivial topology of the magnetic texture.
\begin{figure}[h]
\centering
\includegraphics[width=6.5cm, trim=9 5 5 10, clip]{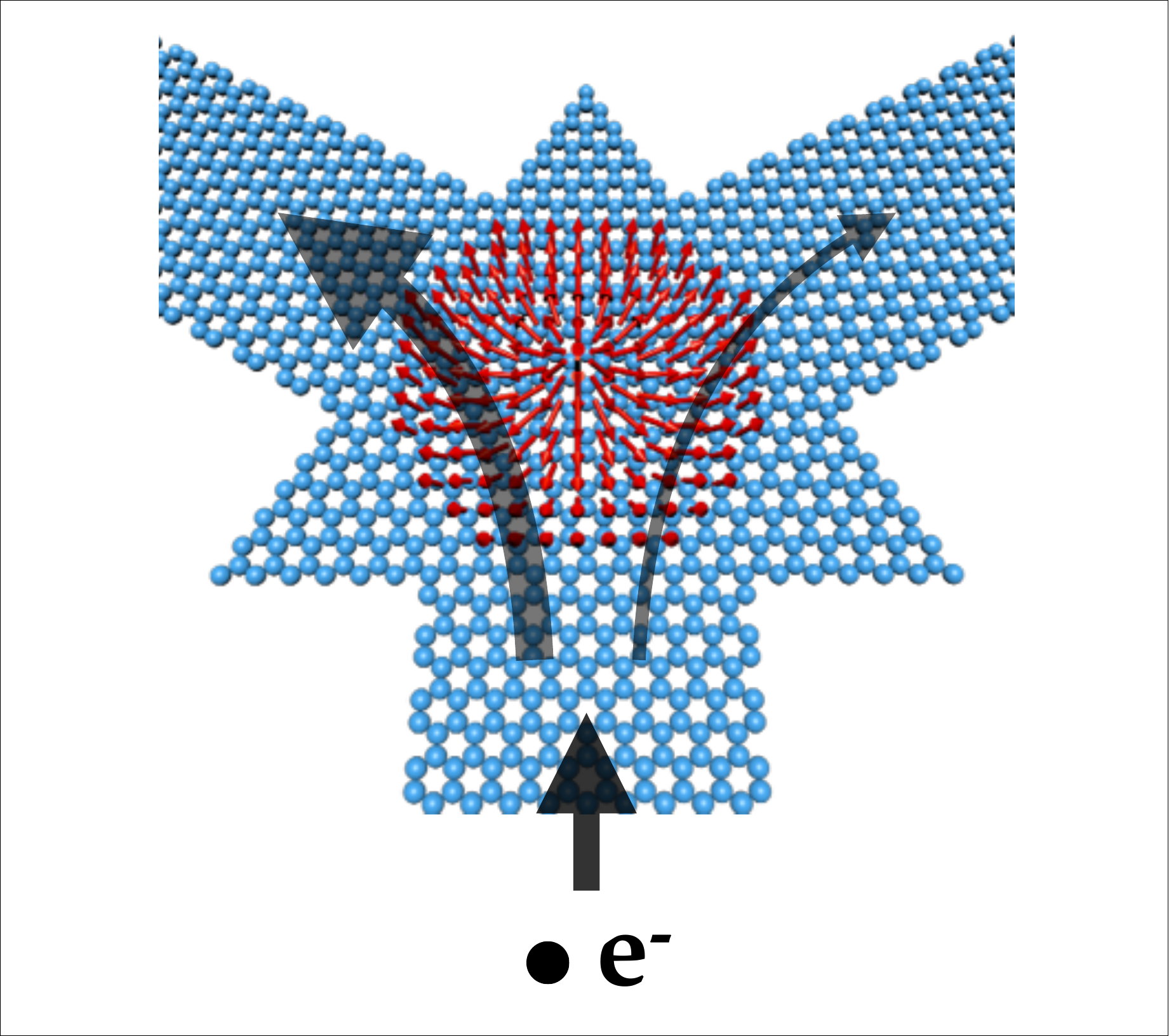}
\caption{A graphene triangular quantum dot (the transmission region) proximized with a skyrmion and connected to three leads. Due to the anomalous Hall effect, a net transverse voltage is generated by the skew scattering of Dirac electrons traveling though the central region.}
\label{fig: trijunction}
\end{figure}

Several recent theoretical works\cite{Hamamoto15,Yin15,Lado15} point out that two-dimensional systems coupled either weakly or strongly to individual skyrmions or skyrmionic lattices can develop an Anomalous Hall (AH) or Quantum Anomalous Hall (QAH) phase owing to the non-trivial topology of these structures in real space. This effect refers to the onset of a transverse Hall response arising in magnetic systems driven by anomalous velocities, associated to Berry curvature, without the need of an applied magnetic field\cite{Nagaosa10}. This anomalous Hall response can be either of extrinsic or intrinsic nature. In the case of proximizing a pristine 2D system with magnetic skyrmions, the generation of a transverse voltage is of extrinsic nature and ascribable to the imprinting of the skyrmions real space topology onto the (trivial) reciprocal space topology of the non-magnetic system\cite{Lado15}.
Based on these findings, along with a recent work demonstrating the possibility
of growing a graphene flake on top of a single atomic layer of Fe on a Ir(111)
substrate\cite{Hamamoto15,Brede14}, here we consider graphene flakes weakly
coupled to magnetic films as skyrmion detectors. To this aim, we
compute the skewness of the scattering and the associated Hall signal induced
in a graphene island coupled to a single skyrmion within a multi-terminal
geometry. Graphene unique properties are ideal to implement the proposed
device. As a fact, being atomically thin maximizes proximity effects, making it an
optimal material to grow on top of magnetic materials. Furthermore, the
fabrication of high quality graphene electronic devices both at the micron
and nanometer scale is absolutely well demonstrated \cite{banszerus2016ballistic,freitag2016electrostatically,shalom2016quantum} 
and its use as a magnetic sensor for magnetic
adsorbates has been already tested experimentally\cite{Candini11,Candini11Nano} and studied theoretically\cite{Gonzalez13}.

The paper is organized as follows. In section \ref{sec: analytic} we discuss
a 2D Dirac system in the continuum coupled to a non-uniform spin texture
and performing a standard rotation in spin space we unveil two types of 
influence on the Dirac electrons. In
section \ref{sec: method} we introduce Landauer's formalism for quantum
transport on the lattice and describe the setup of the proposed Hall experiment.
Finally, in section \ref{sec: results}, we discuss the results obtained by
applying Landauer's formula to a graphene flake coupled to a single skyrmion,
characterizing the Hall conductance as a function of several parameters and
comparing the effectiveness of graphene with that of a standard two-dimensional
electron gas (2DEG).


\section{Analytic approach in the continuum} \label{sec: analytic}
\begin{figure}
\centering
\includegraphics[width=8cm, trim=13 5 5 10, clip]{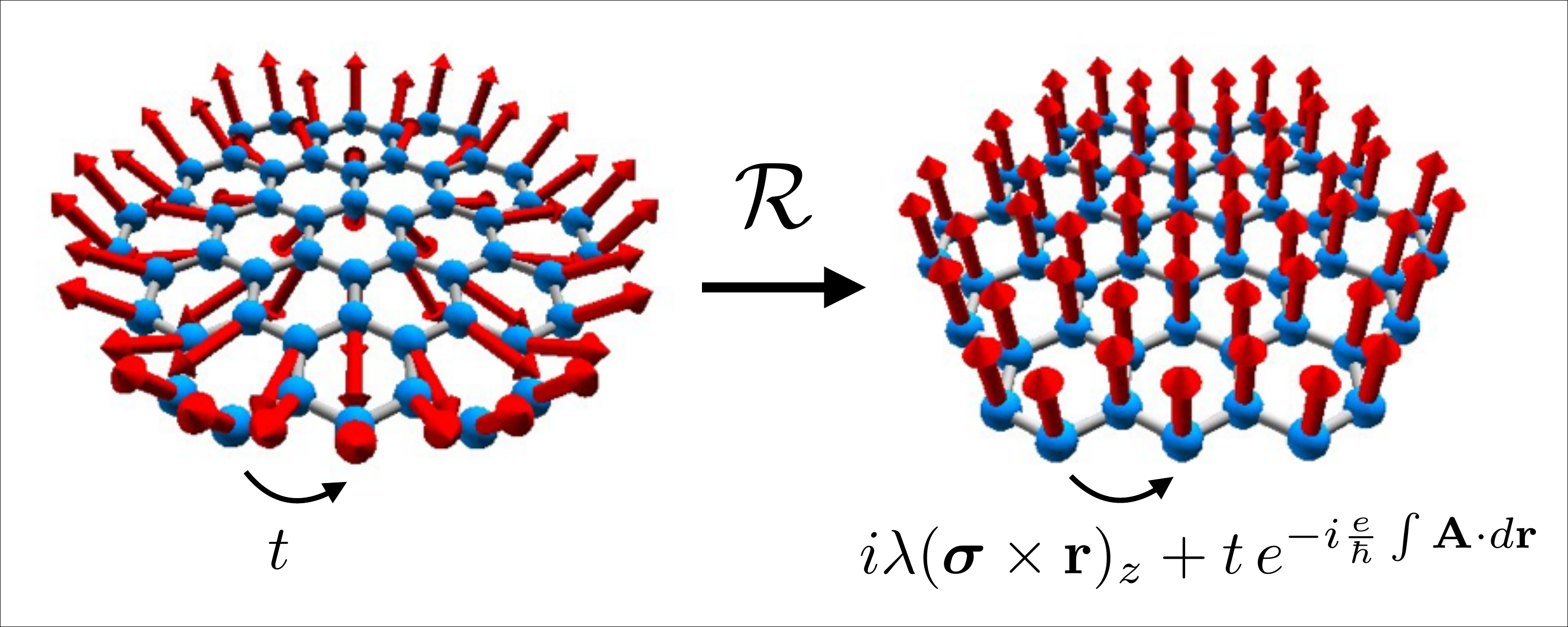}
\caption{Mapping of a system characterized by real hopping and with a double exchange interaction with a non-coplanar magnetic texture to a system with spatially uniform magnetization field and with a complex hopping function mimicking the coexistence of spin-orbit with a vector gauge field.}
\label{fig: mapping}
\end{figure}

In this section we describe graphene electrons interacting with a non-coplanar  magnetization field $\mathbf{n}$,  as given by equation  (\ref{eq: n}),  using a 2D Dirac Hamiltonian:
\begin{equation}
H=H_{0}+H_{ex}=-i\hbar \tau v_{F}\left(\partial_{x}\sigma_{x}+\tau\partial_{y}\sigma_{y}\right)+J\mathbf{n}\cdot\mathbf{s}
\end{equation}
with $\mathbf{s}=(s_{x},s_{y},s_{z})$ the vector of Pauli matrices acting in spin space and $\boldsymbol{\sigma}=(\sigma_{x},\sigma_{y},\sigma_{z})$ the vector of Pauli matrices acting in pseudo-spin space. We perform a rotation of the Hamiltonian so that in every point of space the spin quantization axis is chosen along the direction of the spin texture $\mathbf{n}$. As a result, the representation of the exchange term  is  diagonal in the rotated frame, but the Dirac Hamiltonian acquires new terms that encode the influence of the exchange interaction of the Dirac electrons with the non-coplanar field. The unitary matrix $\mathcal{R}$ that performs such a transformation in the basis $\psi=\left(A\uparrow,B\uparrow,A\downarrow,B\downarrow\right)^{T}$ is
\begin{equation}
\mathcal{R}=
\left(
\begin{array}{cccc}
u & 0 & v & 0\\
0 & u & 0 & v\\
-v^{*} & 0 & u^{*} & 0\\
0 & -v^{*} & 0 & u^{*}
\end{array}\right)
=
\left(
\begin{array}{cc}
u & v\\
-v^* & u^*
\end{array}\right)\otimes\sigma_{0}
\end{equation}
where
\begin{equation}
u=\cos\frac{\Theta(r)}{2}e^{i\Phi(\phi)/2} \quad v=\sin\frac{\Theta(r)}{2}e^{-i\Phi(\phi)/2}
\end{equation}
The transformed Hamiltonian $H \rightarrow H'=\mathcal{R}H\mathcal{R}^{-1}$ reads
\begin{gather}
H'=\tau v_{F}\left[\sigma_{x}\left(p_{x}+\mathcal{A}_{x}\right)+\tau\sigma_{y}\left(p_{y}+\mathcal{A}_{y}\right)\right]+ \nonumber \\
+\frac{\hbar \tau v_{F}}{2}\left[-\sigma_{x}\left(\frac{N}{r}s_{x}n_{y}+\partial_{r}\theta s_{y}\cos\phi\right) \right.+\nonumber \\
+\left. \tau\sigma_{y}\left(\frac{N}{r}s_{x}n_{x}-\partial_{r}\theta s_{y}\sin\phi\right)\right] +Js_{z}
\end{gather}
with
\begin{gather}
\mathcal{A}_{x}=\hbar\frac{N}{2r}\cos\theta\sin\phi\otimes s_{z} \nonumber \\
\mathcal{A}_{y}=-\hbar\frac{N}{2r}\cos\theta\cos\phi\otimes s_{z}
\end{gather}
and $ n_{x}=\cos\Phi\sin\Theta$, $n_{y}=\sin\Phi\sin\Theta $.
In the rotated reference frame, the exchange term is manifestly diagonal. Besides, the Hamiltonian has acquired additional kinetic terms. The $\mathbf{\mathcal{A}}=(\mathcal{A}_{x},\mathcal{A}_{y})$ field acts as a spin-dependent gauge vector potential that couples with the momenta of the Dirac electrons, whereas the remaining two terms closely resemble a spin-orbit (SO) interaction of the Rashba type. On the lattice, this corresponds to mapping a system characterized by a non-collinear exchange field and real hopping to a ferromagnetic system with a purely imaginary hopping mimicking the effect of SO coupling plus a complex hopping supported by a gauge field entering as a Peierls phase. This is schematized in figure \ref{fig: mapping}.
From the gauge field, one can compute the effective magnetic field acting on the system as
\begin{equation}
\mathbf{\mathcal{B}}=\nabla\times\mathbf{\mathcal{A}}=\hbar\frac{N}{2r}s_{z}\left[\partial_{r}\theta\sin\theta-\frac{1}{r}\cos\theta\right]\hat{z}
\end{equation}
that reads
\begin{equation}
\mathcal{B}_z=-\hbar \frac{N}{2r} s_z\left\{ \begin{array}{ll}
\left[\pi \sin\theta / R+r^{-1}\cos\theta\right] & \mbox{for \ensuremath{r\leq R}}\\
r^{-1} & \mbox{for }r>R
\end{array}\right.
\end{equation}
This transformation of the Hamiltonian therefore allows to interpret the topological content embedded in the skyrmion texture as a superposition of two effects: (i) The generation of an effective emergent electromagnetic field (EEMF) described by the gauge potential $\mathbf{\mathcal{A}}$; (ii) The coexistence of ferromagnetic exchange with a Rashba-like SO interaction, what has been predicted to give rise to a QAH phase \cite{Niu10}.
Both ingredients are endowed with a topological character that the skyrmion texture is able to imprint onto the Dirac electrons and are therefore responsible for generating a Hall response in the system. An analog result has been derived\cite{Nagaosa-Tokura13} for Schrodinger electrons, with the remarkable difference that in the strong coupling limit ($J \gg t$) the spin-mixing terms vanish and the problem is exactly mapped to a spinless one-band system where the electrons momenta are coupled to a vector potential describing an emergent magnetic field. In the case of Dirac electrons, the spin-mixing term survives at all coupling regimes and the mapping to a pure EEMF is an incomplete description of the physics taking place in the system. Whereas this picture provides some physical insight of what happens to graphene Dirac electrons surfing a  skyrmions, it does not provide
a straightforward method to compute the Hall response. 


\section{Tight-binding quantum transport approach} \label{sec: method}

In this section we overview the quantum transport  methodology that we will employ to compute the Hall response induced by an individual  magnetic skyrmion in a graphene device. 
Importantly, we are implicitly assuming that the substrate material is an insulating skyrmion crystal such as CuGeO$_3$\cite{Sljivancanin1997} and Cu$_2$OSeO$_3$\cite{Seki12,Langner14,zhang2016imaging} in such a way that the current only flows through graphene.

The graphene electrons are described with the standard tight-binding Hammiltonian for the honeycomb lattice with one $p_z$ orbital per atom\cite{Castro-Neto2009}, plus their exchange interaction with the classical magnetization of the skyrmion $\mathbf{n}$:
\begin{equation}
H=-t\sum_{<i,j>,\sigma}c_{i\sigma}^{\dagger}c_{j\sigma}+J\sum_{i}\textbf{S}_i\cdot\mathbf{n}_{i}
\label{H}
\end{equation}
Here $\textbf{n}_i$ is the classical continuous magnetization texture (\ref{eq: n}) discretized over the graphene lattice and taken at site $i$ and $\textbf{S}_i=\sum_{\sigma\sigma'}c_{i\sigma}^{\dagger} \mathbf{s}_{\sigma\sigma'} c_{i\sigma'}$ is the vector whose components are the Pauli matrices acting in spin space associated with the $i$-th lattice site. The $<i,j>$ symbol implies summation over all nearest neighboring pairs of atoms, and we are assuming that the magnitude of the magnetization is uniform over the whole graphene lattice. This Hamiltonian has been considered before\cite{Lado15} for the case of 2D graphene interacting with a skyrmion crystal. In contrast, here we consider a graphene device that hosts an individual skyrmion.

The mathematical framework that we use to study quantum transport is based on Landauer's formalism for conductance\cite{Landauer57}. Given an experimental setup where a device is attached to $N$ metallic contacts, Landauer's multi-terminal technique allows to compute the transmission amplitude between the $m$-th and the $n$-th contact from the relation
\begin{equation}
T_{mn}=\mbox{Tr}\left(G_{d}^{+}\Gamma_{n}G_{d}\Gamma_{m}\right)
\end{equation}
where $G_{d}$ and $G_{d}^{+}$ are respectively the retarded and advanced
Green's functions of the device, that is the Green's function of the isolated device corrected by the self-energies $\Sigma_m$ of the $N$ leads
\begin{equation}
G_{d}(\epsilon)=\left[\left(\epsilon+i\delta\right)\mathbb{I}-H_{d}-\sum_{m=0}^{N-1}\Sigma_{m}\right]^{-1}
\end{equation}
where $H_{d}$ is the Hamiltonian of the isolated device. The $\Gamma_{m}$'s are quantities associated to the leads' selfenergies as $\Gamma_{m}=i\left(\Sigma_{m}-\Sigma_{m}^{+}\right)$. The leads' self-energies incorporate the coupling between the device and the leads as $\Sigma_{m}=t_{m}^{+}g_{m}t_{m}$, with $g_{m}$ the surface Green's function \cite{sancho1985highly} of the $m$-th lead, and $t_{m}$ the hopping matrix between the device and the $m$-th lead. From the knowledge of the transmission amplitudes, the expression for the total current flowing from the lead $m$ follows straightforwardly:
\begin{equation}
I_{m}=\frac{e}{h}\sum_{n\neq m}\intop_{-\infty}^{+\infty}d\epsilon\left[f\left(\epsilon-\mu_{m}\right)-f\left(\epsilon-\mu_{n}\right)\right]T_{mn}(\epsilon)
\end{equation}
with $f(\epsilon-\mu)$ the Fermi distribution function, so that at zero temperature the previous expression reduces to $ I_{m}=\frac{e}{h}\sum_{n\neq m}\intop_{\mu_{n}}^{\mu_{m}}d\epsilon T_{mn}(\epsilon_F) $ and for a sufficiently small energy interval $\mu_{m}-\mu_{n}$ one can expand the transmission coefficient $T_{mn}\left(\epsilon\right)$ around the Fermi energy $\epsilon_{F}$ and stick to zeroth order. By doing so, one finally finds that the formula for the current flowing from the lead $m$ becomes:
\begin{equation}
I_{m}=\frac{e}{h}\sum_{n\neq m}(\mu_{m}-\mu_{n})T_{mn}(\epsilon_{F})
\label{eq: I}
\end{equation}
This equation can be used to derive the Hall response  in a given  multiterminal device in two different ways. In both cases, the first step of the calculation is the numerical determination of the  transmission coefficients $T_{mn}\left(\epsilon\right)$. Then we can either impose (i) the voltage drops $eV$, defined as the difference between the chemical potentials of the different electrodes, and compute the resulting current (inverse Hall effect), or (ii) impose a longitudinal current flow and a null transverse current, find the resulting chemical potentials and determine the Hall response (direct Hall effect).

\begin{figure}[t]
\centering
\includegraphics[width=7.8cm, trim=13 5 5 10, clip]{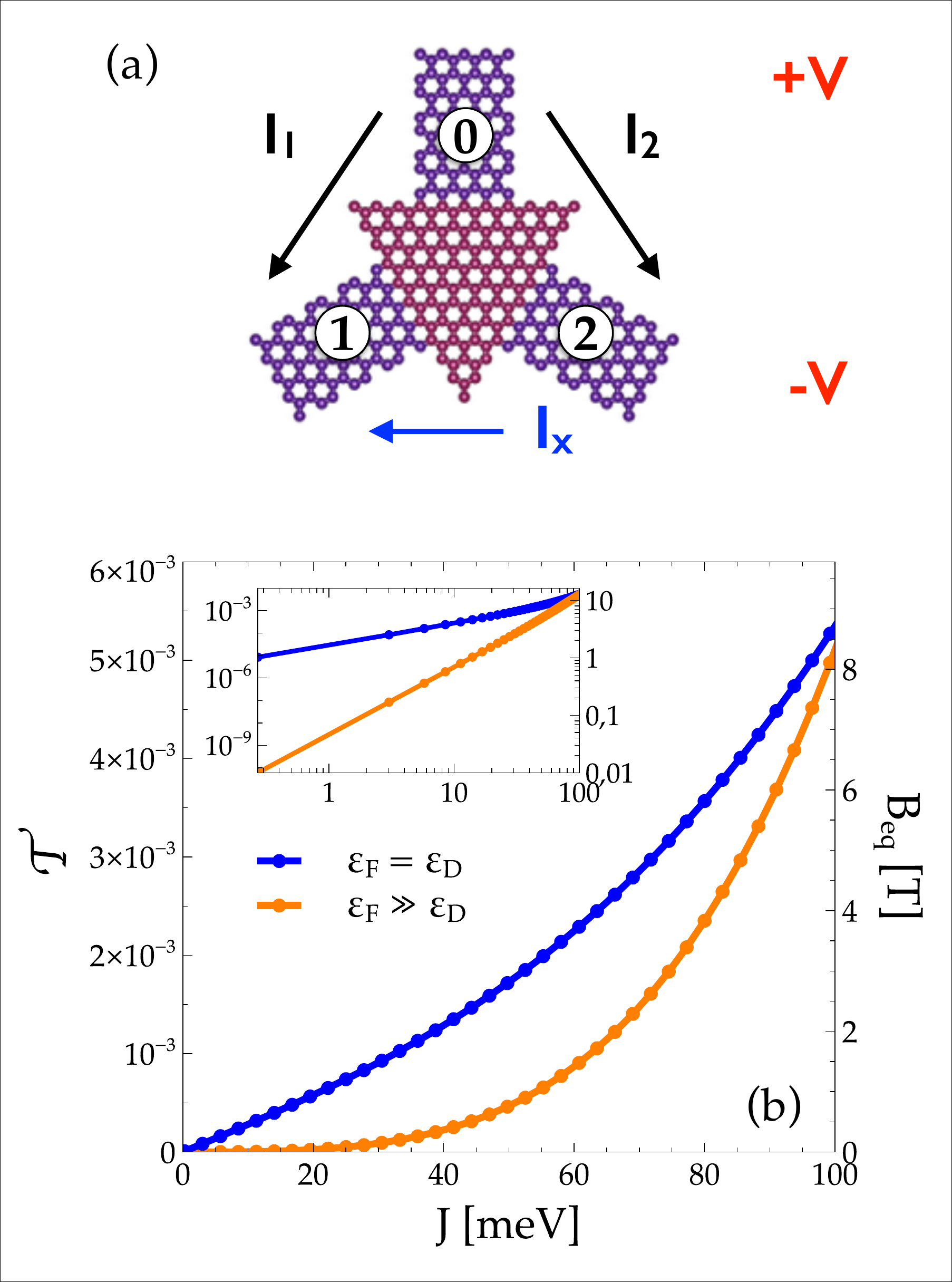}
\caption{(a) Three terminal device setup for the  inverse Hall measurement with $C_3$ rotational symmetry. (b) Normalized transmission imbalance $\mathcal{T}$  (eq. \ref{dT}) and equivalent magnetic field $B_{eq}$ as a function of the coupling constant $J$, comparison of a Dirac-like (undoped graphene) and a Schrodinger-like (heavily doped graphene) material for an island with side of 10.6 nm and a skyrmion radius of 2.3 nm. Inset: log-log representation of $\mathcal{T}(J)$ and $B_{eq}(J)$.}
\label{fig: TTD1}
\end{figure}

When the methods just described are implemented in an ordinary four-terminal geometry\cite{Yin15}, the resulting relation between the Hall conductance and the transmission coefficients is far from intuitive. In this paper, for the sake of simplicity,  we consider a three terminal device (TTD) of the kind of the one shown in figure \ref{fig: TTD1}a.  We choose to fix the chemical potentials of the three electrodes, labeled as $0$, $1$ and $2$, and compute the resulting current.  Specifically, we impose that 
$V_{0}=V$ and $V_{1}=V_{2}=-V$.
In this way, the voltage difference between leads 1 and 2 is automatically set to zero whereas the voltage difference between lead 0 and leads 1,2 is $V_{y}=V_{0}-V_{1,2}=2V$.
The expression for the current flowing from leads 1 and 2 is $I_{i}=2VT_{0i}$ for $ i=1,2$. From this expressions it is straightforward to deduce the current imbalance $\delta I$, that reflects the presence of a transverse 
force, 
$ \delta I=I_{1}-I_{2}=2V(T_{01}-T_{02}) $,
whence our definition of  Hall conductance in this geometry
\begin{equation}
G_{H}=
\frac{\delta I }{V_{0}-V_{1,2}}=\frac{e^2}{h}\left(T_{01}-T_{02}\right)\equiv\frac{e^2}{h} \delta T
\end{equation}
In the following we present the numerical results for the normalized transmission imbalance, that is
\begin{equation}
\mathcal{T}=\frac{\delta T}{T}\equiv (T_{01}-T_{02})/(T_{01}+T_{02})
\label{dT}
\end{equation} in order to work with quantities that do not depend on the number of conduction  channels in the device. This 3-terminal setup simplifies considerably the analysis of the numerical results, and also matches the $C_3$ symmetry of the graphene lattice. However, in a real device, disorder and contact asymmetries might result in additional transmission imbalances that might obscure the detection of skyrmions. Thus, in real devices a standard 4 terminal geometry should be used, given that the principles and magnitude of the physical effect are expected to be the same.



\section{Results and discussion} \label{sec: results}
We now present the results obtained by calculating the imbalance in the transmission coefficients $\mathcal{T}$ eq. (\ref{dT}) for a graphene quantum dot coupled to a skyrmion. For a better physical insight, we provide an estimate for the \textit{equivalent} magnetic field $B_{eq}$ that would give rise to a conventional Hall response of the same  magnitude of that induced by the skyrmion. Details on the determination of such a field are given in the Appendix.
In the following we consider flakes sizes of the order of $\sim 50$
nm$^2$, and skyrmions with radius of the order of 2-3 nm and winding number $N=1$. Also, we are solely
interested in realistic\cite{wei2016strong,Yang13} weak exchange proximity
effects, that do not alter the graphene spectrum substantially, so we explore
coupling constants up to $J\sim 100$ meV\cite{Qiao14}. In order to simulate
standard metallic contacts in some of the calculations square leads have been
used instead of hexagonal leads. Results obtained with different leads
geometries are consistent, so we chose to present curves associated to one or
the other geometry in order to minimize resonance effects due to
confinement inside of the central island.

\subsection{Anomalous Hall effect}
We first investigate the magnitude and behavior of the transmission asymmetry $\mathcal T$ as a function of the coupling constant $J$, comparing the
results for Dirac electrons (half filled honeycomb lattice),
and Schrodinger electrons (heavily doped honeycomb lattice).
The result is shown in fig. \ref{fig: TTD1}(b) in both linear and logarithmic scale,
for a skyrmion with radius $R=2.3$ nm and a device of linear dimension $L=10.6$ nm.
The first thing to notice is that, even for  small $J\simeq 1$ meV, 
the equivalent field $B_{\rm eq}$ is of the order of 1 Tesla,  which shows that the anomalous Hall effect is very large. For $J<100$ meV the transmission imbalance $\mathcal{T}$ of Dirac electrons shows an
approximately linear behavior with $J$ in contrast with the case of Schrodinger electrons (Fermi energy away from the Dirac point)  for which ${\cal T}\propto J^3$. 
For all the values of $J$, the Hall response for Dirac electrons is much larger than for Schrodinger electrons, most notably for the experimentally relevant case of  small $J$, for which $\mathcal{T}$  is up to 4 orders of magnitude larger. This difference is reduced and eventually canceled at higher and unrealistic couplings larger than 100 meV.



\begin{figure}[h]
\centering
\includegraphics[width=8.6cm, trim=13 5 5 10, clip]{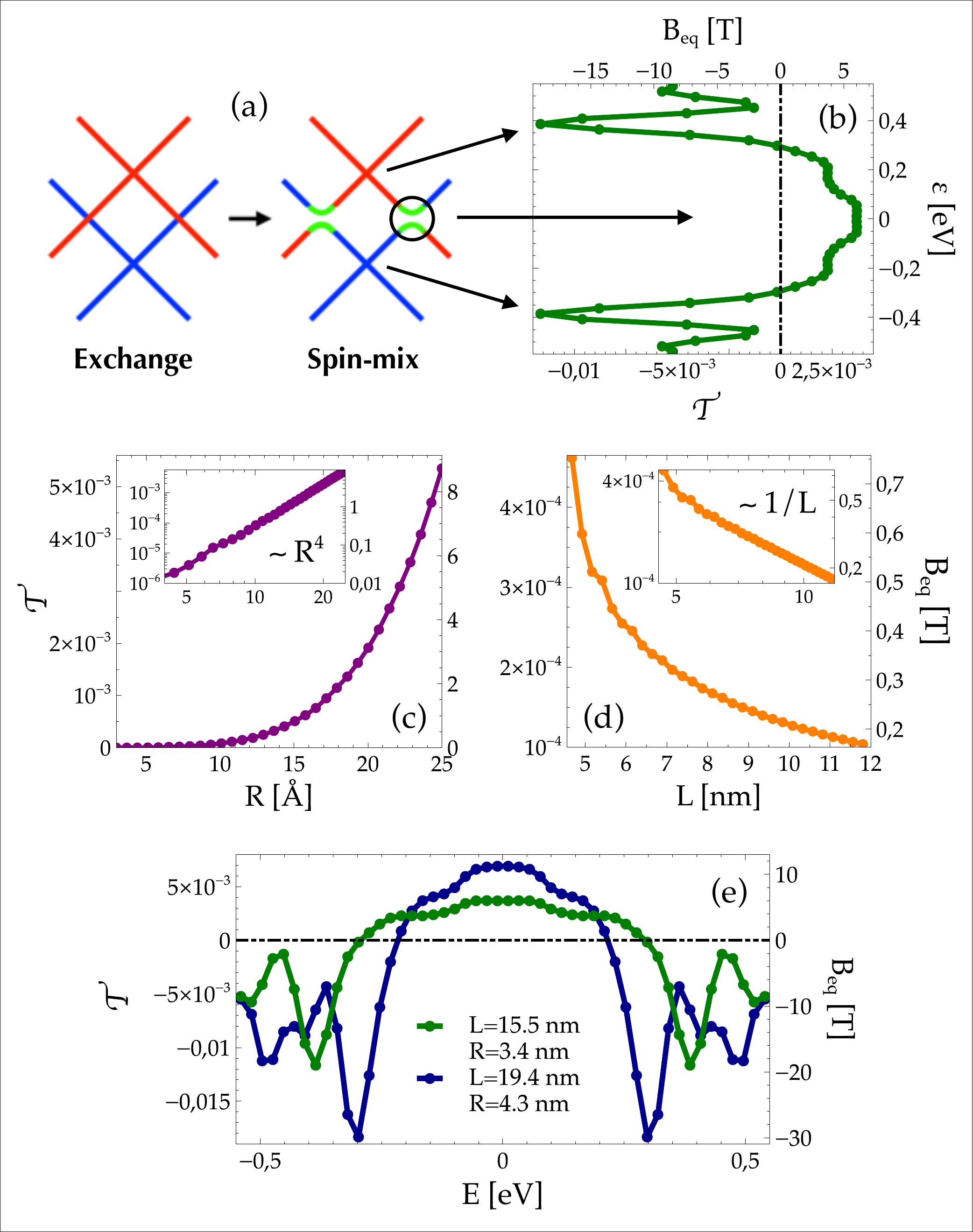}
\caption{(a) Schematics of the effect on the local electronic structure of graphene of being proximized to a skyrmion. (b) Left-right normalized transmission imbalance $\mathcal{T}$ of a graphene TTD as a function of the transmission energy of the leads $\varepsilon$ for an island of 15.5 nm, skyrmion radius of 3.4 nm and coupling constant $J=80$ meV. Energies characterized by maximum absolute Berry curvature in the infinite system are evidenced. (c) and (d) Transmission imbalance of a graphene TTD as a function of skyrmion radius (with fixed flake size of $L=10.6$ nm) and flake size (with fixed skyrmion radius of $R=1.4$ nm), respectively. Both calculations have been performed for a coupling constant of $80$ meV. Insets show log-log representation of $\mathcal{T}$. (e) Comparison of two calculations where the radius of the skyrmion and the linear size of the flake are scaled linearly by a common factor $\alpha=1.25$, for $J=80$ meV. All plots present a second vertical axis in which the equivalent magnetic field $B_{eq}$ is displayed.}
\label{fig: TTD2}
\end{figure}

We now characterize the Hall conductance of a graphene TTD by investigating
its dependence on the system parameters, such as the 
Fermi  energy of the
leads, the skyrmion size $R$  and the  size of the graphene island coupled to the skyrmion. 
The results are shown in
fig. \ref{fig: TTD2}.
The anomalous Hall response as a function of
the chemical potential of graphene (fig. \ref{fig: TTD2}(a,b)),
shows a local maxima at charge neutrality, and other two local maxima of opposite sign at symmetric electron/hole doping,
a behavior resembling graphene coupled to a skyrmion crystal.\cite{Lado15}
Such phenomenology can be understood in terms of
the modification of the Dirac cone due to
the non-coplanar magnetization field. As we have seen in section \ref{sec:
analytic}, the problem can be mapped to one where spatially uniform exchange
field and Rashba-like spin-mixing terms coexist. The
first contribution has the effect of lifting spin degeneracy, whereas the
latter opens small gaps at both the Fermi energy and at crossing points forming
at higher energies of the order of $\pm J$. Within these gaps, the absolute
value of the Berry curvature reaches local maxima and this is reflected in the
behavior of $\mathcal{T}$ as a function of the transmission
energy $\varepsilon$ shown in fig. \ref{fig: TTD2}.

In fig. \ref{fig: TTD2}(c) we show the behavior of $\mathcal{T}$ as a function of the skyrmion
radius $R$, keeping the dimension of the device constant and equal to $L=10.6$ nm, and $J=80$ meV. We consider the case of small skyrmions with nanometric radius such as those found in systems with frustrated exchange
interactions\cite{Okubo2012}. Two competing effects are at play
as the radius of the skyrmion increases: on the one side the change in
magnetization as a function of the distance from the skyrmion center becomes
smoother, so that the effective skew scattering is weaker, and on the other the surface where the skew scattering is non zero increases. The normalized scattering asymmetry resulting from our calculations behaves as $R^4$ indicating that the
second mechanism is dominant, and therefore that larger skyrmions yield a
stronger Hall signal.

The dependence of the Hall response on the size of the graphene flake is shown in Fig. \ref{fig: TTD2}(d), for a fixed radius of $R=1.4$ nm and an exchange of $J=80$ meV. We see that by increasing the flake size while keeping
the skyrmion radius fixed, the Hall signal decreases as $L^{-1}$, where $L$ is the linear size of the triangular transmission region. From these results we infer that the Hall conductance behaves as $\mathcal{T}(R,L)\sim
R^4/L$ as a function of the radius and of the linear size of the central
island. This scaling reflects the fact that the Hall response is proportional to the probability that the electrons surf over the skyrmion, which is manifestly an increasing function of $R$ and a decreasing function of $L$.

By changing both the radius and the device size by a common factor
$\alpha$, $\mathcal{T}$ scales as $\mathcal{T}(\alpha R, \alpha L)\sim \alpha^3 \mathcal{T}(R,L)$
indicating that the Hall conductance is not scale invariant under simultaneous rescaling of $R$ and $L$. Now, since we are
considering flakes of the minimum experimentally achievable dimensions
proximized with the smallest skyrmions experimentally detected so far (of the
order of the nm, whereas observation of skyrmions with radius of up to 100 nm
has been reported\cite{Yu10,Wiesendanger16}), the presented scaling argument evidences that our
estimates of Hall conductances of the order of $10^{-5}$-$10^{-4}$ $G_0$ merely
set a lower bound for the range of values that this parameter can undertake in
actual laboratory measurements. A general example of this non-linear scaling trend is shown
in fig. \ref{fig: TTD2}(e) where a comparison of two systems with $L$ and $R$
scaled by a common factor is presented.

We note that most systems in the brink of hosting skyrmion lattices need a non-zero external magnetic flux to drive them into the skyrmionic phase, as they typically exhibit spiral spin phases at zero magnetic field. This implies that an additional non-zero Hall contribution is to be expected from the external field that sums up to the one driven by the skyrmion alone. An effective way to discriminate between the two effects relies on their different symmetry properties. In fact, while the skyrmionic contribution is electron-hole symmetric (as made clear by fig. \ref{fig: TTD2}(b)) and changes sign only by switching the sign of either $J$ or $N$, the Hall effect induced by the magnetic field is electron-hole asymmetric as holes have opposite charge with respect to electrons and thus respond with an opposite velocity to an applied external magnetic field. It is thus the $\varepsilon\rightarrow -\varepsilon$ asymmetry of the overall scattering cross-section that allows to subtract the spurious external contribution and determine the intrinsic skyrmionic one.

%
%

\subsection{Effects of disorder}
So far we have  dealt with a graphene flake perfectly clean. However, some current degradation brought about by defects or impurities in the sample is to be expected. In order to provide a more realistic estimate of the extent to which the Hall responses that our results anticipate are robust with respect to this loss of conductance, we now consider the effect of  introducing an amount of scalar disorder in the samples.
We do so by averaging over $N=50$ Anderson disorder configurations in each of which we assign a random scalar on-site potential $W_i\,\in\,[-W/2:W/2]$ to each atom in the quantum dot and tune the parameter controlling the disorder degree $W$ from 0 to a maximum of $\sim400$ meV, an upper limit for the energy scale associated with disorder that is consistent with the assumption of Coulomb long-range scattering\cite{Wang15,Nomura2007}. The clean limit is recovered for $W=0$.

\begin{figure}[h]
\centering
\includegraphics[width=8.6cm, trim=13 5 5 10, clip]{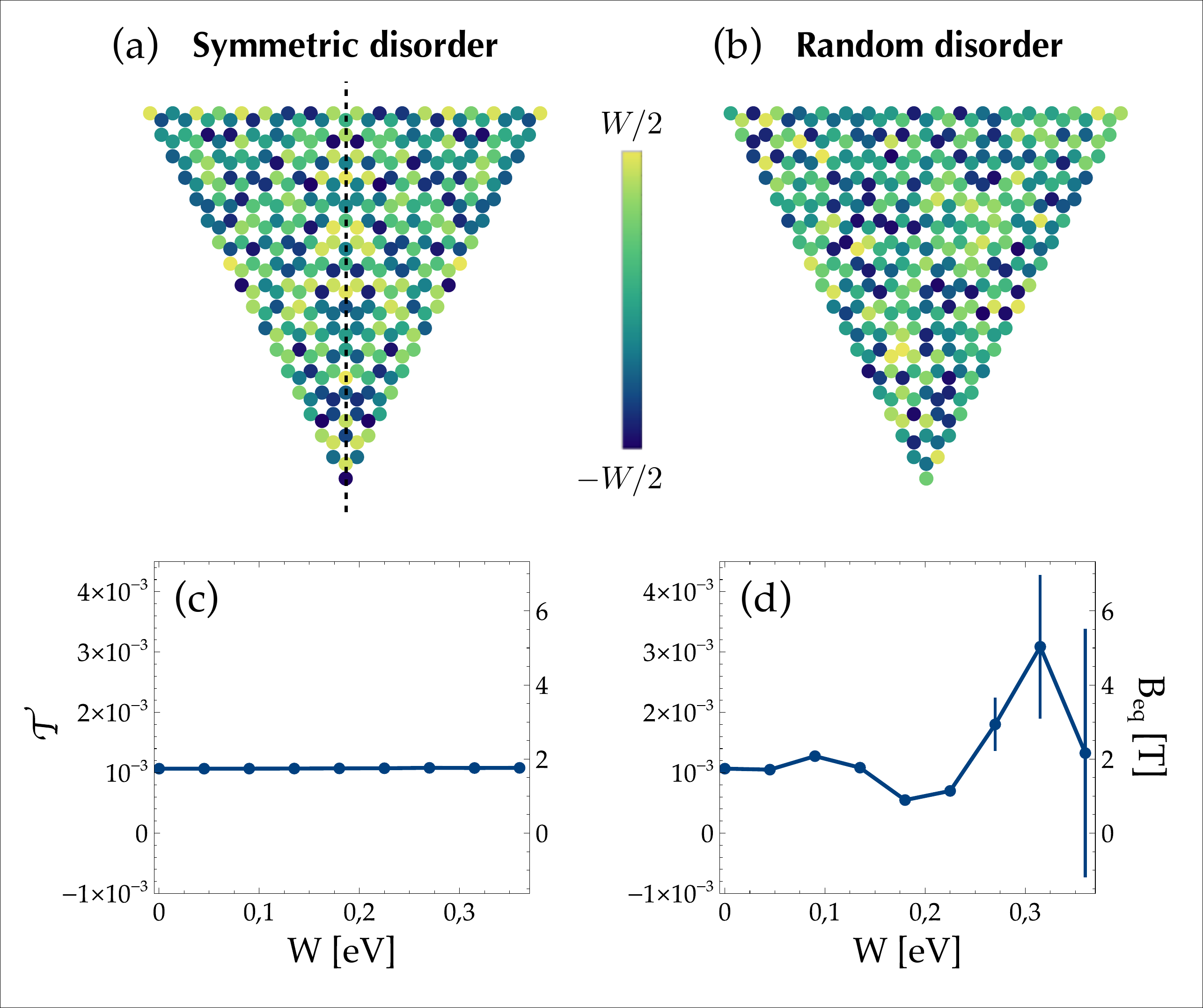}
\caption{Panels (a) and (b) show a typical realization of a disordered
configuration with (a) and without (b) $y\rightarrow -y$ symmetry. In
panels (c) and (d) we present the associated curves of $\mathcal{T}$ and $B_{eq}$ as a function of the disorder strength $W$ for fixed values of $J=80$ meV, $L=10.6$ nm and $R=2.3$ nm.}
\label{fig: Disorder}
\end{figure}

We employ square leads and compare two disorder configurations with different symmetry: one where the
disorder distribution preserves mirror symmetry with respect to the $y$ axis
and one where the distribution is completely random in the whole sample. A realization of each of these different disorder profiles is shown in fig. \ref{fig: Disorder}(a,b). Error bars associated with the standard deviation of the data are shown for completeness.
From the resulting $\mathcal{T}$ curves shown in fig. \ref{fig: Disorder}(c,d) we see that symmetric disorder barely affects the Hall response of the problem, as it provokes changes in the normalized transmission imbalance of the order of $\Delta \mathcal{T}/\mathcal{T}\approx 10^{-2}$. On the other side, a randomly distributed disorder that does not respect $y\rightarrow -y$ symmetry affects the conductance more sizeably, yielding variations $\Delta\mathcal{T}$ of the order of $\mathcal{T}$.
The difference could be explained by noting that in the symmetric case the defects simply act as a fluctuating potential that does not contribute to the asymmetry of the scattering, whereas in the random case an additional transverse conductance driven by the disorder asymmetry rather than by the skyrmion-induced AHE is generated. However, significant alterations of the Hall response only take place at relatively high values of the disorder potential of the order of $\sim 400$ meV,
whereas for weaker and more reasonable disorder strengths the change in the
conductance is smaller and comparable to the one obtained in the symmetric
configuration. We can therefore safely rely on the results obtained so far for pristine graphene, as the unavoidable presence of a low concentration of defects and noise in the actual samples is not able to turn down the figure of merit of the problem.


\section{Conclusions}

Our results strongly indicate that graphene would be an excellent skyrmion
detector at realistic exchange couplings of the order of $\sim 1$-$10$ meV, exhibiting minimum
Hall conductances $G_{H}$ of the order of $10^{-5}$-$10^{-4}$ $G_0$, 
several orders of magnitude larger than the  minimum experimentally detectable
conductance of the order of $10^{-10}$ $G_0$\cite{tzalenchuk2010towards,jeckelmann2001quantum}. The equivalent magnetic field $B_{\rm eq}$ can easily reach one Tesla for $J\approx 1$ meV , $R\approx 2$ nm and $L\approx 10$ nm.
Besides,
these values merely set a lower bound estimate for the
conductances that are detectable in actual experimental devices where sample
dimensions, skyrmion radius and even skyrmion number can be consistently
larger than those considered in this work.
Our results also show that at weak coupling Schrodinger electrons are less sensitive to the non-trivial magnetic ordering and respond with a conductance that is some orders of magnitude smaller than that displayed by Dirac electrons. Finally, we proved that scalar disorder does not affect the transverse conductance in a dramatic manner.

In conclusion, we suggest that graphene might be exploited as a non-invasive probe to readout the presence of an individual skyrmion in a material underneath. The underlying physical principle is the enhanced anomalous Hall effect due to the interaction of Dirac graphene fermions with non-coplanar spin textures. Our work establishes the principles of hybrid devices combining graphene Hall probes and insulating skyrmionic materials\cite{Seki12,Langner14,zhang2016imaging}.

\acknowledgements
The authors acknowledge financial support by Marie-Curie-ITN 607904-SPINOGRAPH. JFR acknowledges financial supported by MEC-Spain (FIS2013-47328-C2-2-P and MAT2016-78625-C2) and Generalitat Valenciana (ACOMP/2010/070), Prometeo, by ERDF funds through the Portuguese Operational Program for Competitiveness and Internationalization COMPETE 2020, and National Funds through FCT- The Portuguese Foundation for Science and Technology, under the project PTDC/FIS-NAN/4662/2014 (016656). This work has been financially supported in part by FEDER funds. JLL and FF thank the hospitality of the Departamento de Fisica Aplicada at the Universidad de Alicante. We are grateful to F. Guinea and P. San-Jose for useful discussions.

\section*{Appendix: Determination of $B_{eq}$}
\begin{figure}[h]
\centering
\includegraphics[width=5.8cm]{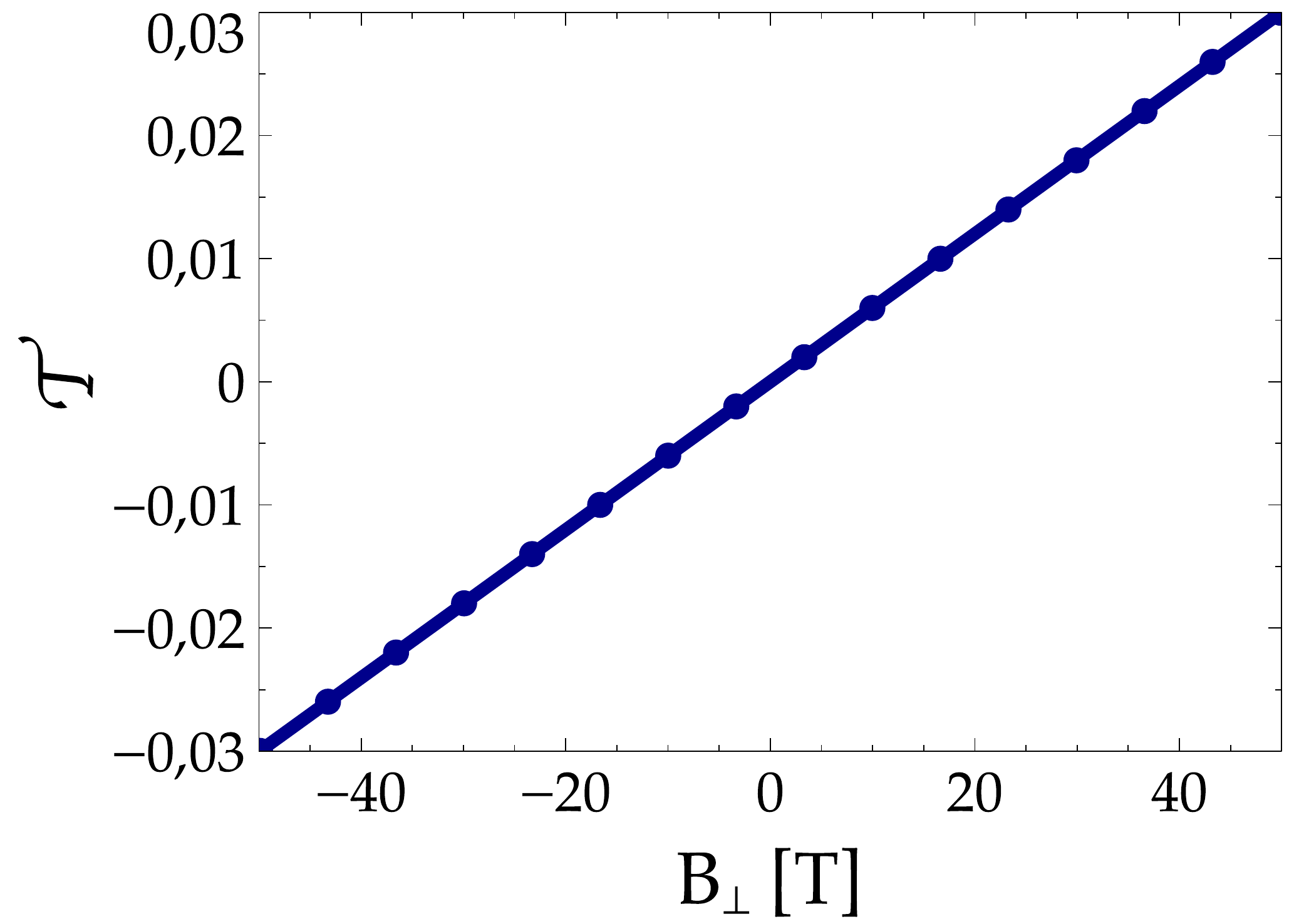}
\caption{Normalized transmission imbalance $\mathcal{T}=\delta T / T$ as a function of an applied perpendicular magnetic field $B_{\perp}$.}
\label{fig:B}
\end{figure}

In order to determine the equivalent magnetic field $B_{eq}$, we have performed a calculation of the transmission imbalance $\mathcal{T}$ of a three-terminal triangular device where a perpendicular magnetic field $B_{\perp}$ is applied to the transmission region. To include such field, we retain only the hopping term of eq. \ref{H} where we perform the standard Peierls substitution $t\rightarrow t\exp{\left(-i\frac{e}{\hbar}\int_{\textbf{r}_i}^{\textbf{r}_j}\textbf{A}\cdot d\textbf{r}\right)}$ such that
\begin{equation}
H=-t\sum_{<i,j>,\sigma} c_{i\sigma}^{\dagger}c_{j\sigma} e^{-i\frac{e}{\hbar}\int_{\textbf{r}_i}^{\textbf{r}_j}\textbf{A}\cdot d\textbf{r}}
\end{equation}
By calculating the transmission imbalance between left and right lead, one gets a linear relation $\mathcal{T}\approx 20 \, B_{\perp}$ as shown in fig. \ref{fig:B}.
The linear relation between $B_{\perp}$ and $\mathcal{T}$, in the absence of a skyrmion, permit to assign an equivalent field $B_{\rm eq}$  to characterize the transmission imbalance calculated in the presence of a skyrmion at $B_{\perp}=0$. 

\bibliography{biblio}

\begin{thebibliography}{45}%
\makeatletter
\providecommand \@ifxundefined [1]{%
 \@ifx{#1\undefined}
}%
\providecommand \@ifnum [1]{%
 \ifnum #1\expandafter \@firstoftwo
 \else \expandafter \@secondoftwo
 \fi
}%
\providecommand \@ifx [1]{%
 \ifx #1\expandafter \@firstoftwo
 \else \expandafter \@secondoftwo
 \fi
}%
\providecommand \natexlab [1]{#1}%
\providecommand \enquote  [1]{``#1''}%
\providecommand \bibnamefont  [1]{#1}%
\providecommand \bibfnamefont [1]{#1}%
\providecommand \citenamefont [1]{#1}%
\providecommand \href@noop [0]{\@secondoftwo}%
\providecommand \href [0]{\begingroup \@sanitize@url \@href}%
\providecommand \@href[1]{\@@startlink{#1}\@@href}%
\providecommand \@@href[1]{\endgroup#1\@@endlink}%
\providecommand \@sanitize@url [0]{\catcode `\\12\catcode `\$12\catcode
  `\&12\catcode `\#12\catcode `\^12\catcode `\_12\catcode `\%12\relax}%
\providecommand \@@startlink[1]{}%
\providecommand \@@endlink[0]{}%
\providecommand \url  [0]{\begingroup\@sanitize@url \@url }%
\providecommand \@url [1]{\endgroup\@href {#1}{\urlprefix }}%
\providecommand \urlprefix  [0]{URL }%
\providecommand \Eprint [0]{\href }%
\providecommand \doibase [0]{http://dx.doi.org/}%
\providecommand \selectlanguage [0]{\@gobble}%
\providecommand \bibinfo  [0]{\@secondoftwo}%
\providecommand \bibfield  [0]{\@secondoftwo}%
\providecommand \translation [1]{[#1]}%
\providecommand \BibitemOpen [0]{}%
\providecommand \bibitemStop [0]{}%
\providecommand \bibitemNoStop [0]{.\EOS\space}%
\providecommand \EOS [0]{\spacefactor3000\relax}%
\providecommand \BibitemShut  [1]{\csname bibitem#1\endcsname}%
\let\auto@bib@innerbib\@empty
\bibitem [{\citenamefont {Roszler}\ \emph {et~al.}(2006)\citenamefont
  {Roszler}, \citenamefont {Bogdanov},\ and\ \citenamefont
  {Pfleiderer}}]{Roszler2006}%
  \BibitemOpen
  \bibfield  {author} {\bibinfo {author} {\bibfnamefont {U.~K.}\ \bibnamefont
  {Roszler}}, \bibinfo {author} {\bibfnamefont {A.~N.}\ \bibnamefont
  {Bogdanov}}, \ and\ \bibinfo {author} {\bibfnamefont {C.}~\bibnamefont
  {Pfleiderer}},\ }\href {http://dx.doi.org/10.1038/nature05056} {\bibfield
  {journal} {\bibinfo  {journal} {Nature}\ }\textbf {\bibinfo {volume} {442}}
  (\bibinfo {year} {2006})}\BibitemShut {NoStop}%
\bibitem [{\citenamefont {Duine}(2013)}]{Duine13}%
  \BibitemOpen
  \bibfield  {author} {\bibinfo {author} {\bibfnamefont {R.}~\bibnamefont
  {Duine}},\ }\href {http://dx.doi.org/10.1038/nnano.2013.233} {\bibfield
  {journal} {\bibinfo  {journal} {Nat Nano}\ }\textbf {\bibinfo {volume} {8}},\
  \bibinfo {pages} {800} (\bibinfo {year} {2013})}\BibitemShut {NoStop}%
\bibitem [{\citenamefont {Rosch}(2017)}]{Rosch2017}%
  \BibitemOpen
  \bibfield  {author} {\bibinfo {author} {\bibfnamefont {A.}~\bibnamefont
  {Rosch}},\ }\href {http://dx.doi.org/10.1038/nnano.2016.244} {\bibfield
  {journal} {\bibinfo  {journal} {Nat Nano}\ }\textbf {\bibinfo {volume}
  {12}},\ \bibinfo {pages} {103} (\bibinfo {year} {2017})}\BibitemShut
  {NoStop}%
\bibitem [{\citenamefont {Dup{\'e}}\ \emph {et~al.}(2016)\citenamefont
  {Dup{\'e}}, \citenamefont {Bihlmayer}, \citenamefont {B{\"o}ttcher},
  \citenamefont {Bl{\"u}gel},\ and\ \citenamefont {Heinze}}]{Dupe16}%
  \BibitemOpen
  \bibfield  {author} {\bibinfo {author} {\bibfnamefont {B.}~\bibnamefont
  {Dup{\'e}}}, \bibinfo {author} {\bibfnamefont {G.}~\bibnamefont {Bihlmayer}},
  \bibinfo {author} {\bibfnamefont {M.}~\bibnamefont {B{\"o}ttcher}}, \bibinfo
  {author} {\bibfnamefont {S.}~\bibnamefont {Bl{\"u}gel}}, \ and\ \bibinfo
  {author} {\bibfnamefont {S.}~\bibnamefont {Heinze}},\ }\href
  {http://dx.doi.org/10.1038/ncomms11779} {\bibfield  {journal} {\bibinfo
  {journal} {Nature Communications}\ }\textbf {\bibinfo {volume} {7}},\
  \bibinfo {pages} {11779 EP } (\bibinfo {year} {2016})}\BibitemShut {NoStop}%
\bibitem [{Edi(2013)}]{Editorial2013}%
  \BibitemOpen
  \href {http://dx.doi.org/10.1038/nnano.2013.282} {\bibfield  {journal}
  {\bibinfo  {journal} {Nat Nano}\ }\textbf {\bibinfo {volume} {8}},\ \bibinfo
  {pages} {883} (\bibinfo {year} {2013})}\BibitemShut {NoStop}%
\bibitem [{\citenamefont {M{\"u}hlbauer}\ \emph {et~al.}(2009)\citenamefont
  {M{\"u}hlbauer}, \citenamefont {Binz}, \citenamefont {Jonietz}, \citenamefont
  {Pfleiderer}, \citenamefont {Rosch}, \citenamefont {Neubauer}, \citenamefont
  {Georgii},\ and\ \citenamefont {B{\"o}ni}}]{Muhlbauer09}%
  \BibitemOpen
  \bibfield  {author} {\bibinfo {author} {\bibfnamefont {S.}~\bibnamefont
  {M{\"u}hlbauer}}, \bibinfo {author} {\bibfnamefont {B.}~\bibnamefont {Binz}},
  \bibinfo {author} {\bibfnamefont {F.}~\bibnamefont {Jonietz}}, \bibinfo
  {author} {\bibfnamefont {C.}~\bibnamefont {Pfleiderer}}, \bibinfo {author}
  {\bibfnamefont {A.}~\bibnamefont {Rosch}}, \bibinfo {author} {\bibfnamefont
  {A.}~\bibnamefont {Neubauer}}, \bibinfo {author} {\bibfnamefont
  {R.}~\bibnamefont {Georgii}}, \ and\ \bibinfo {author} {\bibfnamefont
  {P.}~\bibnamefont {B{\"o}ni}},\ }\href@noop {} {\bibfield  {journal}
  {\bibinfo  {journal} {Science}\ }\textbf {\bibinfo {volume} {323}},\ \bibinfo
  {pages} {915} (\bibinfo {year} {2009})}\BibitemShut {NoStop}%
\bibitem [{\citenamefont {M\"unzer}\ \emph {et~al.}(2010)\citenamefont
  {M\"unzer}, \citenamefont {Neubauer}, \citenamefont {Adams}, \citenamefont
  {M\"uhlbauer}, \citenamefont {Franz}, \citenamefont {Jonietz}, \citenamefont
  {Georgii}, \citenamefont {B\"oni}, \citenamefont {Pedersen}, \citenamefont
  {Schmidt}, \citenamefont {Rosch},\ and\ \citenamefont
  {Pfleiderer}}]{Munzer10}%
  \BibitemOpen
  \bibfield  {author} {\bibinfo {author} {\bibfnamefont {W.}~\bibnamefont
  {M\"unzer}}, \bibinfo {author} {\bibfnamefont {A.}~\bibnamefont {Neubauer}},
  \bibinfo {author} {\bibfnamefont {T.}~\bibnamefont {Adams}}, \bibinfo
  {author} {\bibfnamefont {S.}~\bibnamefont {M\"uhlbauer}}, \bibinfo {author}
  {\bibfnamefont {C.}~\bibnamefont {Franz}}, \bibinfo {author} {\bibfnamefont
  {F.}~\bibnamefont {Jonietz}}, \bibinfo {author} {\bibfnamefont
  {R.}~\bibnamefont {Georgii}}, \bibinfo {author} {\bibfnamefont
  {P.}~\bibnamefont {B\"oni}}, \bibinfo {author} {\bibfnamefont
  {B.}~\bibnamefont {Pedersen}}, \bibinfo {author} {\bibfnamefont
  {M.}~\bibnamefont {Schmidt}}, \bibinfo {author} {\bibfnamefont
  {A.}~\bibnamefont {Rosch}}, \ and\ \bibinfo {author} {\bibfnamefont
  {C.}~\bibnamefont {Pfleiderer}},\ }\href {\doibase
  10.1103/PhysRevB.81.041203} {\bibfield  {journal} {\bibinfo  {journal} {Phys.
  Rev. B}\ }\textbf {\bibinfo {volume} {81}},\ \bibinfo {pages} {041203}
  (\bibinfo {year} {2010})}\BibitemShut {NoStop}%
\bibitem [{\citenamefont {Yu}\ \emph {et~al.}(2012)\citenamefont {Yu},
  \citenamefont {Kanazawa}, \citenamefont {Zhang}, \citenamefont {Nagai},
  \citenamefont {Hara}, \citenamefont {Kimoto}, \citenamefont {Matsui},
  \citenamefont {Onose},\ and\ \citenamefont {Tokura}}]{yu2012skyrmion}%
  \BibitemOpen
  \bibfield  {author} {\bibinfo {author} {\bibfnamefont {X.}~\bibnamefont
  {Yu}}, \bibinfo {author} {\bibfnamefont {N.}~\bibnamefont {Kanazawa}},
  \bibinfo {author} {\bibfnamefont {W.}~\bibnamefont {Zhang}}, \bibinfo
  {author} {\bibfnamefont {T.}~\bibnamefont {Nagai}}, \bibinfo {author}
  {\bibfnamefont {T.}~\bibnamefont {Hara}}, \bibinfo {author} {\bibfnamefont
  {K.}~\bibnamefont {Kimoto}}, \bibinfo {author} {\bibfnamefont
  {Y.}~\bibnamefont {Matsui}}, \bibinfo {author} {\bibfnamefont
  {Y.}~\bibnamefont {Onose}}, \ and\ \bibinfo {author} {\bibfnamefont
  {Y.}~\bibnamefont {Tokura}},\ }\href@noop {} {\bibfield  {journal} {\bibinfo
  {journal} {Nature communications}\ }\textbf {\bibinfo {volume} {3}},\
  \bibinfo {pages} {988} (\bibinfo {year} {2012})}\BibitemShut {NoStop}%
\bibitem [{\citenamefont {Tokunaga}\ \emph {et~al.}(2015)\citenamefont
  {Tokunaga}, \citenamefont {Yu}, \citenamefont {White}, \citenamefont
  {R{\o}nnow}, \citenamefont {Morikawa}, \citenamefont {Taguchi},\ and\
  \citenamefont {Tokura}}]{tokunaga2015new}%
  \BibitemOpen
  \bibfield  {author} {\bibinfo {author} {\bibfnamefont {Y.}~\bibnamefont
  {Tokunaga}}, \bibinfo {author} {\bibfnamefont {X.}~\bibnamefont {Yu}},
  \bibinfo {author} {\bibfnamefont {J.}~\bibnamefont {White}}, \bibinfo
  {author} {\bibfnamefont {H.~M.}\ \bibnamefont {R{\o}nnow}}, \bibinfo {author}
  {\bibfnamefont {D.}~\bibnamefont {Morikawa}}, \bibinfo {author}
  {\bibfnamefont {Y.}~\bibnamefont {Taguchi}}, \ and\ \bibinfo {author}
  {\bibfnamefont {Y.}~\bibnamefont {Tokura}},\ }\href@noop {} {\bibfield
  {journal} {\bibinfo  {journal} {Nature communications}\ }\textbf {\bibinfo
  {volume} {6}} (\bibinfo {year} {2015})}\BibitemShut {NoStop}%
\bibitem [{\citenamefont {Seki}\ \emph {et~al.}(2012)\citenamefont {Seki},
  \citenamefont {Yu}, \citenamefont {Ishiwata},\ and\ \citenamefont
  {Tokura}}]{Seki12}%
  \BibitemOpen
  \bibfield  {author} {\bibinfo {author} {\bibfnamefont {S.}~\bibnamefont
  {Seki}}, \bibinfo {author} {\bibfnamefont {X.}~\bibnamefont {Yu}}, \bibinfo
  {author} {\bibfnamefont {S.}~\bibnamefont {Ishiwata}}, \ and\ \bibinfo
  {author} {\bibfnamefont {Y.}~\bibnamefont {Tokura}},\ }\href@noop {}
  {\bibfield  {journal} {\bibinfo  {journal} {Science}\ }\textbf {\bibinfo
  {volume} {336}},\ \bibinfo {pages} {198} (\bibinfo {year}
  {2012})}\BibitemShut {NoStop}%
\bibitem [{\citenamefont {Langner}\ \emph {et~al.}(2014)\citenamefont
  {Langner}, \citenamefont {Roy}, \citenamefont {Mishra}, \citenamefont {Lee},
  \citenamefont {Shi}, \citenamefont {Hossain}, \citenamefont {Chuang},
  \citenamefont {Seki}, \citenamefont {Tokura}, \citenamefont {Kevan},\ and\
  \citenamefont {Schoenlein}}]{Langner14}%
  \BibitemOpen
  \bibfield  {author} {\bibinfo {author} {\bibfnamefont {M.~C.}\ \bibnamefont
  {Langner}}, \bibinfo {author} {\bibfnamefont {S.}~\bibnamefont {Roy}},
  \bibinfo {author} {\bibfnamefont {S.~K.}\ \bibnamefont {Mishra}}, \bibinfo
  {author} {\bibfnamefont {J.~C.~T.}\ \bibnamefont {Lee}}, \bibinfo {author}
  {\bibfnamefont {X.~W.}\ \bibnamefont {Shi}}, \bibinfo {author} {\bibfnamefont
  {M.~A.}\ \bibnamefont {Hossain}}, \bibinfo {author} {\bibfnamefont {Y.-D.}\
  \bibnamefont {Chuang}}, \bibinfo {author} {\bibfnamefont {S.}~\bibnamefont
  {Seki}}, \bibinfo {author} {\bibfnamefont {Y.}~\bibnamefont {Tokura}},
  \bibinfo {author} {\bibfnamefont {S.~D.}\ \bibnamefont {Kevan}}, \ and\
  \bibinfo {author} {\bibfnamefont {R.~W.}\ \bibnamefont {Schoenlein}},\ }\href
  {\doibase 10.1103/PhysRevLett.112.167202} {\bibfield  {journal} {\bibinfo
  {journal} {Phys. Rev. Lett.}\ }\textbf {\bibinfo {volume} {112}},\ \bibinfo
  {pages} {167202} (\bibinfo {year} {2014})}\BibitemShut {NoStop}%
\bibitem [{\citenamefont {Zhang}\ \emph {et~al.}(2016)\citenamefont {Zhang},
  \citenamefont {Bauer}, \citenamefont {Berger}, \citenamefont {Pfleiderer},
  \citenamefont {van~der Laan},\ and\ \citenamefont
  {Hesjedal}}]{zhang2016imaging}%
  \BibitemOpen
  \bibfield  {author} {\bibinfo {author} {\bibfnamefont {S.}~\bibnamefont
  {Zhang}}, \bibinfo {author} {\bibfnamefont {A.}~\bibnamefont {Bauer}},
  \bibinfo {author} {\bibfnamefont {H.}~\bibnamefont {Berger}}, \bibinfo
  {author} {\bibfnamefont {C.}~\bibnamefont {Pfleiderer}}, \bibinfo {author}
  {\bibfnamefont {G.}~\bibnamefont {van~der Laan}}, \ and\ \bibinfo {author}
  {\bibfnamefont {T.}~\bibnamefont {Hesjedal}},\ }\href@noop {} {\bibfield
  {journal} {\bibinfo  {journal} {Applied Physics Letters}\ }\textbf {\bibinfo
  {volume} {109}},\ \bibinfo {pages} {192406} (\bibinfo {year}
  {2016})}\BibitemShut {NoStop}%
\bibitem [{\citenamefont {Heinze}\ \emph {et~al.}(2011)\citenamefont {Heinze},
  \citenamefont {von Bergmann}, \citenamefont {Menzel}, \citenamefont {Brede},
  \citenamefont {Kubetzka}, \citenamefont {Wiesendanger}, \citenamefont
  {Bihlmayer},\ and\ \citenamefont {Blugel}}]{Heinze11}%
  \BibitemOpen
  \bibfield  {author} {\bibinfo {author} {\bibfnamefont {S.}~\bibnamefont
  {Heinze}}, \bibinfo {author} {\bibfnamefont {K.}~\bibnamefont {von
  Bergmann}}, \bibinfo {author} {\bibfnamefont {M.}~\bibnamefont {Menzel}},
  \bibinfo {author} {\bibfnamefont {J.}~\bibnamefont {Brede}}, \bibinfo
  {author} {\bibfnamefont {A.}~\bibnamefont {Kubetzka}}, \bibinfo {author}
  {\bibfnamefont {R.}~\bibnamefont {Wiesendanger}}, \bibinfo {author}
  {\bibfnamefont {G.}~\bibnamefont {Bihlmayer}}, \ and\ \bibinfo {author}
  {\bibfnamefont {S.}~\bibnamefont {Blugel}},\ }\href
  {http://dx.doi.org/10.1038/nphys2045} {\bibfield  {journal} {\bibinfo
  {journal} {Nat Phys}\ }\textbf {\bibinfo {volume} {7}},\ \bibinfo {pages}
  {713} (\bibinfo {year} {2011})}\BibitemShut {NoStop}%
\bibitem [{\citenamefont {Romming}\ \emph {et~al.}(2015)\citenamefont
  {Romming}, \citenamefont {Kubetzka}, \citenamefont {Hanneken}, \citenamefont
  {von Bergmann},\ and\ \citenamefont {Wiesendanger}}]{Romming15}%
  \BibitemOpen
  \bibfield  {author} {\bibinfo {author} {\bibfnamefont {N.}~\bibnamefont
  {Romming}}, \bibinfo {author} {\bibfnamefont {A.}~\bibnamefont {Kubetzka}},
  \bibinfo {author} {\bibfnamefont {C.}~\bibnamefont {Hanneken}}, \bibinfo
  {author} {\bibfnamefont {K.}~\bibnamefont {von Bergmann}}, \ and\ \bibinfo
  {author} {\bibfnamefont {R.}~\bibnamefont {Wiesendanger}},\ }\href {\doibase
  10.1103/PhysRevLett.114.177203} {\bibfield  {journal} {\bibinfo  {journal}
  {Phys. Rev. Lett.}\ }\textbf {\bibinfo {volume} {114}},\ \bibinfo {pages}
  {177203} (\bibinfo {year} {2015})}\BibitemShut {NoStop}%
\bibitem [{\citenamefont {Yu}\ \emph {et~al.}(2010)\citenamefont {Yu},
  \citenamefont {Onose}, \citenamefont {Kanazawa}, \citenamefont {Park},
  \citenamefont {Han}, \citenamefont {Matsui}, \citenamefont {Nagaosa},\ and\
  \citenamefont {Tokura}}]{Yu10}%
  \BibitemOpen
  \bibfield  {author} {\bibinfo {author} {\bibfnamefont {X.~Z.}\ \bibnamefont
  {Yu}}, \bibinfo {author} {\bibfnamefont {Y.}~\bibnamefont {Onose}}, \bibinfo
  {author} {\bibfnamefont {N.}~\bibnamefont {Kanazawa}}, \bibinfo {author}
  {\bibfnamefont {J.~H.}\ \bibnamefont {Park}}, \bibinfo {author}
  {\bibfnamefont {J.~H.}\ \bibnamefont {Han}}, \bibinfo {author} {\bibfnamefont
  {Y.}~\bibnamefont {Matsui}}, \bibinfo {author} {\bibfnamefont
  {N.}~\bibnamefont {Nagaosa}}, \ and\ \bibinfo {author} {\bibfnamefont
  {Y.}~\bibnamefont {Tokura}},\ }\href {http://dx.doi.org/10.1038/nature09124}
  {\bibfield  {journal} {\bibinfo  {journal} {Nature}\ }\textbf {\bibinfo
  {volume} {465}},\ \bibinfo {pages} {901} (\bibinfo {year}
  {2010})}\BibitemShut {NoStop}%
\bibitem [{\citenamefont {Pfleiderer}(2011)}]{Pfleiderer11}%
  \BibitemOpen
  \bibfield  {author} {\bibinfo {author} {\bibfnamefont {C.}~\bibnamefont
  {Pfleiderer}},\ }\href {http://dx.doi.org/10.1038/nphys2081} {\bibfield
  {journal} {\bibinfo  {journal} {Nat Phys}\ }\textbf {\bibinfo {volume} {7}},\
  \bibinfo {pages} {673} (\bibinfo {year} {2011})}\BibitemShut {NoStop}%
\bibitem [{\citenamefont {Dovzhenko}\ \emph {et~al.}(2016)\citenamefont
  {Dovzhenko}, \citenamefont {Casola}, \citenamefont {Schlotter}, \citenamefont
  {Zhou}, \citenamefont {B{\"u}ttner}, \citenamefont {Walsworth}, \citenamefont
  {Beach},\ and\ \citenamefont {Yacoby}}]{dovzhenko2016imaging}%
  \BibitemOpen
  \bibfield  {author} {\bibinfo {author} {\bibfnamefont {Y.}~\bibnamefont
  {Dovzhenko}}, \bibinfo {author} {\bibfnamefont {F.}~\bibnamefont {Casola}},
  \bibinfo {author} {\bibfnamefont {S.}~\bibnamefont {Schlotter}}, \bibinfo
  {author} {\bibfnamefont {T.~X.}\ \bibnamefont {Zhou}}, \bibinfo {author}
  {\bibfnamefont {F.}~\bibnamefont {B{\"u}ttner}}, \bibinfo {author}
  {\bibfnamefont {R.~L.}\ \bibnamefont {Walsworth}}, \bibinfo {author}
  {\bibfnamefont {G.~S.}\ \bibnamefont {Beach}}, \ and\ \bibinfo {author}
  {\bibfnamefont {A.}~\bibnamefont {Yacoby}},\ }\href@noop {} {\bibfield
  {journal} {\bibinfo  {journal} {arXiv preprint arXiv:1611.00673}\ } (\bibinfo
  {year} {2016})}\BibitemShut {NoStop}%
\bibitem [{\citenamefont {Parkin}\ \emph {et~al.}(2008)\citenamefont {Parkin},
  \citenamefont {Hayashi},\ and\ \citenamefont {Thomas}}]{Parkin08}%
  \BibitemOpen
  \bibfield  {author} {\bibinfo {author} {\bibfnamefont {S.~S.}\ \bibnamefont
  {Parkin}}, \bibinfo {author} {\bibfnamefont {M.}~\bibnamefont {Hayashi}}, \
  and\ \bibinfo {author} {\bibfnamefont {L.}~\bibnamefont {Thomas}},\
  }\href@noop {} {\bibfield  {journal} {\bibinfo  {journal} {Science}\ }\textbf
  {\bibinfo {volume} {320}},\ \bibinfo {pages} {190} (\bibinfo {year}
  {2008})}\BibitemShut {NoStop}%
\bibitem [{\citenamefont {Jonietz}\ \emph {et~al.}(2010)\citenamefont
  {Jonietz}, \citenamefont {M{\"u}hlbauer}, \citenamefont {Pfleiderer},
  \citenamefont {Neubauer}, \citenamefont {M{\"u}nzer}, \citenamefont {Bauer},
  \citenamefont {Adams}, \citenamefont {Georgii}, \citenamefont {B{\"o}ni},
  \citenamefont {Duine}, \citenamefont {Everschor}, \citenamefont {Garst},\
  and\ \citenamefont {Rosch}}]{Jonietz10}%
  \BibitemOpen
  \bibfield  {author} {\bibinfo {author} {\bibfnamefont {F.}~\bibnamefont
  {Jonietz}}, \bibinfo {author} {\bibfnamefont {S.}~\bibnamefont
  {M{\"u}hlbauer}}, \bibinfo {author} {\bibfnamefont {C.}~\bibnamefont
  {Pfleiderer}}, \bibinfo {author} {\bibfnamefont {A.}~\bibnamefont
  {Neubauer}}, \bibinfo {author} {\bibfnamefont {W.}~\bibnamefont
  {M{\"u}nzer}}, \bibinfo {author} {\bibfnamefont {A.}~\bibnamefont {Bauer}},
  \bibinfo {author} {\bibfnamefont {T.}~\bibnamefont {Adams}}, \bibinfo
  {author} {\bibfnamefont {R.}~\bibnamefont {Georgii}}, \bibinfo {author}
  {\bibfnamefont {P.}~\bibnamefont {B{\"o}ni}}, \bibinfo {author}
  {\bibfnamefont {R.~A.}\ \bibnamefont {Duine}}, \bibinfo {author}
  {\bibfnamefont {K.}~\bibnamefont {Everschor}}, \bibinfo {author}
  {\bibfnamefont {M.}~\bibnamefont {Garst}}, \ and\ \bibinfo {author}
  {\bibfnamefont {A.}~\bibnamefont {Rosch}},\ }\href
  {http://science.sciencemag.org/content/330/6011/1648.abstract} {\bibfield
  {journal} {\bibinfo  {journal} {Science}\ }\textbf {\bibinfo {volume}
  {330}},\ \bibinfo {pages} {1648} (\bibinfo {year} {2010})}\BibitemShut
  {NoStop}%
\bibitem [{\citenamefont {Nagaosa}\ and\ \citenamefont
  {Tokura}(2013)}]{Nagaosa2013}%
  \BibitemOpen
  \bibfield  {author} {\bibinfo {author} {\bibfnamefont {N.}~\bibnamefont
  {Nagaosa}}\ and\ \bibinfo {author} {\bibfnamefont {Y.}~\bibnamefont
  {Tokura}},\ }\href {http://dx.doi.org/10.1038/nnano.2013.243} {\bibfield
  {journal} {\bibinfo  {journal} {Nat Nano}\ }\textbf {\bibinfo {volume} {8}},\
  \bibinfo {pages} {899} (\bibinfo {year} {2013})}\BibitemShut {NoStop}%
\bibitem [{\citenamefont {Hamamoto}\ \emph {et~al.}(2015)\citenamefont
  {Hamamoto}, \citenamefont {Ezawa},\ and\ \citenamefont
  {Nagaosa}}]{Hamamoto15}%
  \BibitemOpen
  \bibfield  {author} {\bibinfo {author} {\bibfnamefont {K.}~\bibnamefont
  {Hamamoto}}, \bibinfo {author} {\bibfnamefont {M.}~\bibnamefont {Ezawa}}, \
  and\ \bibinfo {author} {\bibfnamefont {N.}~\bibnamefont {Nagaosa}},\ }\href
  {http://link.aps.org/doi/10.1103/PhysRevB.92.115417} {\bibfield  {journal}
  {\bibinfo  {journal} {Physical Review B}\ }\textbf {\bibinfo {volume} {92}},\
  \bibinfo {pages} {115417} (\bibinfo {year} {2015})}\BibitemShut {NoStop}%
\bibitem [{\citenamefont {Yin}\ \emph {et~al.}(2015)\citenamefont {Yin},
  \citenamefont {Liu}, \citenamefont {Barlas}, \citenamefont {Zang},\ and\
  \citenamefont {Lake}}]{Yin15}%
  \BibitemOpen
  \bibfield  {author} {\bibinfo {author} {\bibfnamefont {G.}~\bibnamefont
  {Yin}}, \bibinfo {author} {\bibfnamefont {Y.}~\bibnamefont {Liu}}, \bibinfo
  {author} {\bibfnamefont {Y.}~\bibnamefont {Barlas}}, \bibinfo {author}
  {\bibfnamefont {J.}~\bibnamefont {Zang}}, \ and\ \bibinfo {author}
  {\bibfnamefont {R.~K.}\ \bibnamefont {Lake}},\ }\href
  {http://link.aps.org/doi/10.1103/PhysRevB.92.024411} {\bibfield  {journal}
  {\bibinfo  {journal} {Physical Review B}\ }\textbf {\bibinfo {volume} {92}},\
  \bibinfo {pages} {024411} (\bibinfo {year} {2015})}\BibitemShut {NoStop}%
\bibitem [{\citenamefont {Lado}\ and\ \citenamefont
  {Fern\'andez-Rossier}(2015)}]{Lado15}%
  \BibitemOpen
  \bibfield  {author} {\bibinfo {author} {\bibfnamefont {J.~L.}\ \bibnamefont
  {Lado}}\ and\ \bibinfo {author} {\bibfnamefont {J.}~\bibnamefont
  {Fern\'andez-Rossier}},\ }\href {\doibase 10.1103/PhysRevB.92.115433}
  {\bibfield  {journal} {\bibinfo  {journal} {Phys. Rev. B}\ }\textbf {\bibinfo
  {volume} {92}},\ \bibinfo {pages} {115433} (\bibinfo {year}
  {2015})}\BibitemShut {NoStop}%
\bibitem [{\citenamefont {Nagaosa}\ \emph {et~al.}(2010)\citenamefont
  {Nagaosa}, \citenamefont {Sinova}, \citenamefont {Onoda}, \citenamefont
  {MacDonald},\ and\ \citenamefont {Ong}}]{Nagaosa10}%
  \BibitemOpen
  \bibfield  {author} {\bibinfo {author} {\bibfnamefont {N.}~\bibnamefont
  {Nagaosa}}, \bibinfo {author} {\bibfnamefont {J.}~\bibnamefont {Sinova}},
  \bibinfo {author} {\bibfnamefont {S.}~\bibnamefont {Onoda}}, \bibinfo
  {author} {\bibfnamefont {A.~H.}\ \bibnamefont {MacDonald}}, \ and\ \bibinfo
  {author} {\bibfnamefont {N.~P.}\ \bibnamefont {Ong}},\ }\href
  {http://link.aps.org/doi/10.1103/RevModPhys.82.1539} {\bibfield  {journal}
  {\bibinfo  {journal} {Reviews of Modern Physics}\ }\textbf {\bibinfo {volume}
  {82}},\ \bibinfo {pages} {1539} (\bibinfo {year} {2010})}\BibitemShut
  {NoStop}%
\bibitem [{\citenamefont {Brede}\ \emph {et~al.}(2014)\citenamefont {Brede},
  \citenamefont {Atodiresei}, \citenamefont {Caciuc}, \citenamefont {Bazarnik},
  \citenamefont {Al-Zubi}, \citenamefont {Bl{\"u}gel},\ and\ \citenamefont
  {Wiesendanger}}]{Brede14}%
  \BibitemOpen
  \bibfield  {author} {\bibinfo {author} {\bibfnamefont {J.}~\bibnamefont
  {Brede}}, \bibinfo {author} {\bibfnamefont {N.}~\bibnamefont {Atodiresei}},
  \bibinfo {author} {\bibfnamefont {V.}~\bibnamefont {Caciuc}}, \bibinfo
  {author} {\bibfnamefont {M.}~\bibnamefont {Bazarnik}}, \bibinfo {author}
  {\bibfnamefont {A.}~\bibnamefont {Al-Zubi}}, \bibinfo {author} {\bibfnamefont
  {S.}~\bibnamefont {Bl{\"u}gel}}, \ and\ \bibinfo {author} {\bibfnamefont
  {R.}~\bibnamefont {Wiesendanger}},\ }\href@noop {} {\bibfield  {journal}
  {\bibinfo  {journal} {Nature nanotechnology}\ }\textbf {\bibinfo {volume}
  {9}},\ \bibinfo {pages} {1018} (\bibinfo {year} {2014})}\BibitemShut
  {NoStop}%
\bibitem [{\citenamefont {Banszerus}\ \emph {et~al.}(2016)\citenamefont
  {Banszerus}, \citenamefont {Schmitz}, \citenamefont {Engels}, \citenamefont
  {Goldsche}, \citenamefont {Watanabe}, \citenamefont {Taniguchi},
  \citenamefont {Beschoten},\ and\ \citenamefont
  {Stampfer}}]{banszerus2016ballistic}%
  \BibitemOpen
  \bibfield  {author} {\bibinfo {author} {\bibfnamefont {L.}~\bibnamefont
  {Banszerus}}, \bibinfo {author} {\bibfnamefont {M.}~\bibnamefont {Schmitz}},
  \bibinfo {author} {\bibfnamefont {S.}~\bibnamefont {Engels}}, \bibinfo
  {author} {\bibfnamefont {M.}~\bibnamefont {Goldsche}}, \bibinfo {author}
  {\bibfnamefont {K.}~\bibnamefont {Watanabe}}, \bibinfo {author}
  {\bibfnamefont {T.}~\bibnamefont {Taniguchi}}, \bibinfo {author}
  {\bibfnamefont {B.}~\bibnamefont {Beschoten}}, \ and\ \bibinfo {author}
  {\bibfnamefont {C.}~\bibnamefont {Stampfer}},\ }\href@noop {} {\bibfield
  {journal} {\bibinfo  {journal} {Nano letters}\ }\textbf {\bibinfo {volume}
  {16}},\ \bibinfo {pages} {1387} (\bibinfo {year} {2016})}\BibitemShut
  {NoStop}%
\bibitem [{\citenamefont {Freitag}\ \emph {et~al.}(2016)\citenamefont
  {Freitag}, \citenamefont {Chizhova}, \citenamefont {Nemes-Incze},
  \citenamefont {Woods}, \citenamefont {Gorbachev}, \citenamefont {Cao},
  \citenamefont {Geim}, \citenamefont {Novoselov}, \citenamefont {Burgdorfer},
  \citenamefont {Libisch} \emph {et~al.}}]{freitag2016electrostatically}%
  \BibitemOpen
  \bibfield  {author} {\bibinfo {author} {\bibfnamefont {N.~M.}\ \bibnamefont
  {Freitag}}, \bibinfo {author} {\bibfnamefont {L.~A.}\ \bibnamefont
  {Chizhova}}, \bibinfo {author} {\bibfnamefont {P.}~\bibnamefont
  {Nemes-Incze}}, \bibinfo {author} {\bibfnamefont {C.~R.}\ \bibnamefont
  {Woods}}, \bibinfo {author} {\bibfnamefont {R.~V.}\ \bibnamefont
  {Gorbachev}}, \bibinfo {author} {\bibfnamefont {Y.}~\bibnamefont {Cao}},
  \bibinfo {author} {\bibfnamefont {A.~K.}\ \bibnamefont {Geim}}, \bibinfo
  {author} {\bibfnamefont {K.~S.}\ \bibnamefont {Novoselov}}, \bibinfo {author}
  {\bibfnamefont {J.}~\bibnamefont {Burgdorfer}}, \bibinfo {author}
  {\bibfnamefont {F.}~\bibnamefont {Libisch}},  \emph {et~al.},\ }\href@noop {}
  {\bibfield  {journal} {\bibinfo  {journal} {Nano Letters}\ }\textbf {\bibinfo
  {volume} {16}},\ \bibinfo {pages} {5798} (\bibinfo {year}
  {2016})}\BibitemShut {NoStop}%
\bibitem [{\citenamefont {Shalom}\ \emph {et~al.}(2016)\citenamefont {Shalom},
  \citenamefont {Zhu}, \citenamefont {Falko}, \citenamefont {Mishchenko},
  \citenamefont {Kretinin}, \citenamefont {Novoselov}, \citenamefont {Woods},
  \citenamefont {Watanabe}, \citenamefont {Taniguchi}, \citenamefont {Geim}
  \emph {et~al.}}]{shalom2016quantum}%
  \BibitemOpen
  \bibfield  {author} {\bibinfo {author} {\bibfnamefont {M.~B.}\ \bibnamefont
  {Shalom}}, \bibinfo {author} {\bibfnamefont {M.}~\bibnamefont {Zhu}},
  \bibinfo {author} {\bibfnamefont {V.}~\bibnamefont {Falko}}, \bibinfo
  {author} {\bibfnamefont {A.}~\bibnamefont {Mishchenko}}, \bibinfo {author}
  {\bibfnamefont {A.}~\bibnamefont {Kretinin}}, \bibinfo {author}
  {\bibfnamefont {K.}~\bibnamefont {Novoselov}}, \bibinfo {author}
  {\bibfnamefont {C.}~\bibnamefont {Woods}}, \bibinfo {author} {\bibfnamefont
  {K.}~\bibnamefont {Watanabe}}, \bibinfo {author} {\bibfnamefont
  {T.}~\bibnamefont {Taniguchi}}, \bibinfo {author} {\bibfnamefont
  {A.}~\bibnamefont {Geim}},  \emph {et~al.},\ }\href@noop {} {\bibfield
  {journal} {\bibinfo  {journal} {Nature Physics}\ }\textbf {\bibinfo {volume}
  {12}},\ \bibinfo {pages} {318} (\bibinfo {year} {2016})}\BibitemShut
  {NoStop}%
\bibitem [{\citenamefont {Candini}\ \emph
  {et~al.}(2011{\natexlab{a}})\citenamefont {Candini}, \citenamefont {Alvino},
  \citenamefont {Wernsdorfer},\ and\ \citenamefont {Affronte}}]{Candini11}%
  \BibitemOpen
  \bibfield  {author} {\bibinfo {author} {\bibfnamefont {A.}~\bibnamefont
  {Candini}}, \bibinfo {author} {\bibfnamefont {C.}~\bibnamefont {Alvino}},
  \bibinfo {author} {\bibfnamefont {W.}~\bibnamefont {Wernsdorfer}}, \ and\
  \bibinfo {author} {\bibfnamefont {M.}~\bibnamefont {Affronte}},\ }\href
  {\doibase 10.1103/PhysRevB.83.121401} {\bibfield  {journal} {\bibinfo
  {journal} {Phys. Rev. B}\ }\textbf {\bibinfo {volume} {83}},\ \bibinfo
  {pages} {121401} (\bibinfo {year} {2011}{\natexlab{a}})}\BibitemShut
  {NoStop}%
\bibitem [{\citenamefont {Candini}\ \emph
  {et~al.}(2011{\natexlab{b}})\citenamefont {Candini}, \citenamefont
  {Klyatskaya}, \citenamefont {Ruben}, \citenamefont {Wernsdorfer},\ and\
  \citenamefont {Affronte}}]{Candini11Nano}%
  \BibitemOpen
  \bibfield  {author} {\bibinfo {author} {\bibfnamefont {A.}~\bibnamefont
  {Candini}}, \bibinfo {author} {\bibfnamefont {S.}~\bibnamefont {Klyatskaya}},
  \bibinfo {author} {\bibfnamefont {M.}~\bibnamefont {Ruben}}, \bibinfo
  {author} {\bibfnamefont {W.}~\bibnamefont {Wernsdorfer}}, \ and\ \bibinfo
  {author} {\bibfnamefont {M.}~\bibnamefont {Affronte}},\ }\bibfield
  {booktitle} {\emph {\bibinfo {booktitle} {Nano Letters}},\ }\href {\doibase
  10.1021/nl2006142} {\bibfield  {journal} {\bibinfo  {journal} {Nano Letters}\
  }\textbf {\bibinfo {volume} {11}},\ \bibinfo {pages} {2634} (\bibinfo {year}
  {2011}{\natexlab{b}})}\BibitemShut {NoStop}%
\bibitem [{\citenamefont {Gonz\'alez}\ \emph {et~al.}(2013)\citenamefont
  {Gonz\'alez}, \citenamefont {Delgado},\ and\ \citenamefont
  {Fern\'andez-Rossier}}]{Gonzalez13}%
  \BibitemOpen
  \bibfield  {author} {\bibinfo {author} {\bibfnamefont {J.~W.}\ \bibnamefont
  {Gonz\'alez}}, \bibinfo {author} {\bibfnamefont {F.}~\bibnamefont {Delgado}},
  \ and\ \bibinfo {author} {\bibfnamefont {J.}~\bibnamefont
  {Fern\'andez-Rossier}},\ }\href {\doibase 10.1103/PhysRevB.87.085433}
  {\bibfield  {journal} {\bibinfo  {journal} {Phys. Rev. B}\ }\textbf {\bibinfo
  {volume} {87}},\ \bibinfo {pages} {085433} (\bibinfo {year}
  {2013})}\BibitemShut {NoStop}%
\bibitem [{\citenamefont {Qiao}\ \emph {et~al.}(2010)\citenamefont {Qiao},
  \citenamefont {Yang}, \citenamefont {Feng}, \citenamefont {Tse},
  \citenamefont {Ding}, \citenamefont {Yao}, \citenamefont {Wang},\ and\
  \citenamefont {Niu}}]{Niu10}%
  \BibitemOpen
  \bibfield  {author} {\bibinfo {author} {\bibfnamefont {Z.}~\bibnamefont
  {Qiao}}, \bibinfo {author} {\bibfnamefont {S.~A.}\ \bibnamefont {Yang}},
  \bibinfo {author} {\bibfnamefont {W.}~\bibnamefont {Feng}}, \bibinfo {author}
  {\bibfnamefont {W.-K.}\ \bibnamefont {Tse}}, \bibinfo {author} {\bibfnamefont
  {J.}~\bibnamefont {Ding}}, \bibinfo {author} {\bibfnamefont {Y.}~\bibnamefont
  {Yao}}, \bibinfo {author} {\bibfnamefont {J.}~\bibnamefont {Wang}}, \ and\
  \bibinfo {author} {\bibfnamefont {Q.}~\bibnamefont {Niu}},\ }\href {\doibase
  10.1103/PhysRevB.82.161414} {\bibfield  {journal} {\bibinfo  {journal} {Phys.
  Rev. B}\ }\textbf {\bibinfo {volume} {82}},\ \bibinfo {pages} {161414}
  (\bibinfo {year} {2010})}\BibitemShut {NoStop}%
\bibitem [{\citenamefont {{\v S}ljivan{\v c}anin}\ \emph
  {et~al.}(1997)\citenamefont {{\v S}ljivan{\v c}anin}, \citenamefont
  {Popovi{\'c}},\ and\ \citenamefont {Vukajlovi{\'c}}}]{Sljivancanin1997}%
  \BibitemOpen
  \bibfield  {author} {\bibinfo {author} {\bibfnamefont {{\v Z}.~V.}\
  \bibnamefont {{\v S}ljivan{\v c}anin}}, \bibinfo {author} {\bibfnamefont
  {Z.~S.}\ \bibnamefont {Popovi{\'c}}}, \ and\ \bibinfo {author} {\bibfnamefont
  {F.~R.}\ \bibnamefont {Vukajlovi{\'c}}},\ }\href
  {http://link.aps.org/doi/10.1103/PhysRevB.56.4432} {\bibfield  {journal}
  {\bibinfo  {journal} {Physical Review B}\ }\textbf {\bibinfo {volume} {56}},\
  \bibinfo {pages} {4432} (\bibinfo {year} {1997})}\BibitemShut {NoStop}%
\bibitem [{\citenamefont {Castro~Neto}\ \emph {et~al.}(2009)\citenamefont
  {Castro~Neto}, \citenamefont {Guinea}, \citenamefont {Peres}, \citenamefont
  {Novoselov},\ and\ \citenamefont {Geim}}]{Castro-Neto2009}%
  \BibitemOpen
  \bibfield  {author} {\bibinfo {author} {\bibfnamefont {A.~H.}\ \bibnamefont
  {Castro~Neto}}, \bibinfo {author} {\bibfnamefont {F.}~\bibnamefont {Guinea}},
  \bibinfo {author} {\bibfnamefont {N.~M.~R.}\ \bibnamefont {Peres}}, \bibinfo
  {author} {\bibfnamefont {K.~S.}\ \bibnamefont {Novoselov}}, \ and\ \bibinfo
  {author} {\bibfnamefont {A.~K.}\ \bibnamefont {Geim}},\ }\href
  {http://link.aps.org/doi/10.1103/RevModPhys.81.109} {\bibfield  {journal}
  {\bibinfo  {journal} {Reviews of Modern Physics}\ }\textbf {\bibinfo {volume}
  {81}},\ \bibinfo {pages} {109} (\bibinfo {year} {2009})}\BibitemShut
  {NoStop}%
\bibitem [{\citenamefont {Landauer}(1957)}]{Landauer57}%
  \BibitemOpen
  \bibfield  {author} {\bibinfo {author} {\bibfnamefont {R.}~\bibnamefont
  {Landauer}},\ }\bibfield  {booktitle} {\emph {\bibinfo {booktitle} {IBM
  Journal of Research and Development}},\ }\href {\doibase 10.1147/rd.13.0223}
  {\bibfield  {journal} {\bibinfo  {journal} {IBM Journal of Research and
  Development}\ }\textbf {\bibinfo {volume} {1}},\ \bibinfo {pages} {223}
  (\bibinfo {year} {1957})}\BibitemShut {NoStop}%
\bibitem [{\citenamefont {Sancho}\ \emph {et~al.}(1985)\citenamefont {Sancho},
  \citenamefont {Sancho}, \citenamefont {Sancho},\ and\ \citenamefont
  {Rubio}}]{sancho1985highly}%
  \BibitemOpen
  \bibfield  {author} {\bibinfo {author} {\bibfnamefont {M.~L.}\ \bibnamefont
  {Sancho}}, \bibinfo {author} {\bibfnamefont {J.~L.}\ \bibnamefont {Sancho}},
  \bibinfo {author} {\bibfnamefont {J.~L.}\ \bibnamefont {Sancho}}, \ and\
  \bibinfo {author} {\bibfnamefont {J.}~\bibnamefont {Rubio}},\ }\href@noop {}
  {\bibfield  {journal} {\bibinfo  {journal} {Journal of Physics F: Metal
  Physics}\ }\textbf {\bibinfo {volume} {15}},\ \bibinfo {pages} {851}
  (\bibinfo {year} {1985})}\BibitemShut {NoStop}%
\bibitem [{\citenamefont {Wei}\ \emph {et~al.}(2016)\citenamefont {Wei},
  \citenamefont {Lee}, \citenamefont {Lemaitre}, \citenamefont {Pinel},
  \citenamefont {Cutaia}, \citenamefont {Cha}, \citenamefont {Katmis},
  \citenamefont {Zhu}, \citenamefont {Heiman}, \citenamefont {Hone} \emph
  {et~al.}}]{wei2016strong}%
  \BibitemOpen
  \bibfield  {author} {\bibinfo {author} {\bibfnamefont {P.}~\bibnamefont
  {Wei}}, \bibinfo {author} {\bibfnamefont {S.}~\bibnamefont {Lee}}, \bibinfo
  {author} {\bibfnamefont {F.}~\bibnamefont {Lemaitre}}, \bibinfo {author}
  {\bibfnamefont {L.}~\bibnamefont {Pinel}}, \bibinfo {author} {\bibfnamefont
  {D.}~\bibnamefont {Cutaia}}, \bibinfo {author} {\bibfnamefont
  {W.}~\bibnamefont {Cha}}, \bibinfo {author} {\bibfnamefont {F.}~\bibnamefont
  {Katmis}}, \bibinfo {author} {\bibfnamefont {Y.}~\bibnamefont {Zhu}},
  \bibinfo {author} {\bibfnamefont {D.}~\bibnamefont {Heiman}}, \bibinfo
  {author} {\bibfnamefont {J.}~\bibnamefont {Hone}},  \emph {et~al.},\
  }\href@noop {} {\bibfield  {journal} {\bibinfo  {journal} {Nature materials}\
  } (\bibinfo {year} {2016})}\BibitemShut {NoStop}%
\bibitem [{\citenamefont {Yang}\ \emph {et~al.}(2013)\citenamefont {Yang},
  \citenamefont {Hallal}, \citenamefont {Terrade}, \citenamefont {Waintal},
  \citenamefont {Roche},\ and\ \citenamefont {Chshiev}}]{Yang13}%
  \BibitemOpen
  \bibfield  {author} {\bibinfo {author} {\bibfnamefont {H.~X.}\ \bibnamefont
  {Yang}}, \bibinfo {author} {\bibfnamefont {A.}~\bibnamefont {Hallal}},
  \bibinfo {author} {\bibfnamefont {D.}~\bibnamefont {Terrade}}, \bibinfo
  {author} {\bibfnamefont {X.}~\bibnamefont {Waintal}}, \bibinfo {author}
  {\bibfnamefont {S.}~\bibnamefont {Roche}}, \ and\ \bibinfo {author}
  {\bibfnamefont {M.}~\bibnamefont {Chshiev}},\ }\href {\doibase
  10.1103/PhysRevLett.110.046603} {\bibfield  {journal} {\bibinfo  {journal}
  {Phys. Rev. Lett.}\ }\textbf {\bibinfo {volume} {110}},\ \bibinfo {pages}
  {046603} (\bibinfo {year} {2013})}\BibitemShut {NoStop}%
\bibitem [{\citenamefont {Qiao}\ \emph {et~al.}(2014)\citenamefont {Qiao},
  \citenamefont {Ren}, \citenamefont {Chen}, \citenamefont {Bellaiche},
  \citenamefont {Zhang}, \citenamefont {MacDonald},\ and\ \citenamefont
  {Niu}}]{Qiao14}%
  \BibitemOpen
  \bibfield  {author} {\bibinfo {author} {\bibfnamefont {Z.}~\bibnamefont
  {Qiao}}, \bibinfo {author} {\bibfnamefont {W.}~\bibnamefont {Ren}}, \bibinfo
  {author} {\bibfnamefont {H.}~\bibnamefont {Chen}}, \bibinfo {author}
  {\bibfnamefont {L.}~\bibnamefont {Bellaiche}}, \bibinfo {author}
  {\bibfnamefont {Z.}~\bibnamefont {Zhang}}, \bibinfo {author} {\bibfnamefont
  {A.~H.}\ \bibnamefont {MacDonald}}, \ and\ \bibinfo {author} {\bibfnamefont
  {Q.}~\bibnamefont {Niu}},\ }\href
  {http://link.aps.org/doi/10.1103/PhysRevLett.112.116404} {\bibfield
  {journal} {\bibinfo  {journal} {Physical Review Letters}\ }\textbf {\bibinfo
  {volume} {112}},\ \bibinfo {pages} {116404} (\bibinfo {year}
  {2014})}\BibitemShut {NoStop}%
\bibitem [{\citenamefont {Okubo}\ \emph {et~al.}(2012)\citenamefont {Okubo},
  \citenamefont {Chung},\ and\ \citenamefont {Kawamura}}]{Okubo2012}%
  \BibitemOpen
  \bibfield  {author} {\bibinfo {author} {\bibfnamefont {T.}~\bibnamefont
  {Okubo}}, \bibinfo {author} {\bibfnamefont {S.}~\bibnamefont {Chung}}, \ and\
  \bibinfo {author} {\bibfnamefont {H.}~\bibnamefont {Kawamura}},\ }\href
  {http://link.aps.org/doi/10.1103/PhysRevLett.108.017206} {\bibfield
  {journal} {\bibinfo  {journal} {Physical Review Letters}\ }\textbf {\bibinfo
  {volume} {108}},\ \bibinfo {pages} {017206} (\bibinfo {year}
  {2012})}\BibitemShut {NoStop}%
\bibitem [{\citenamefont {Wiesendanger}(2016)}]{Wiesendanger16}%
  \BibitemOpen
  \bibfield  {author} {\bibinfo {author} {\bibfnamefont {R.}~\bibnamefont
  {Wiesendanger}},\ }\href {http://dx.doi.org/10.1038/natrevmats.2016.44}
  {\bibfield  {journal} {\bibinfo  {journal} {Nature Reviews Materials}\
  }\textbf {\bibinfo {volume} {1}},\ \bibinfo {pages} {16044 EP } (\bibinfo
  {year} {2016})}\BibitemShut {NoStop}%
\bibitem [{\citenamefont {Wang}\ \emph {et~al.}(2015)\citenamefont {Wang},
  \citenamefont {Tang}, \citenamefont {Sachs}, \citenamefont {Barlas},\ and\
  \citenamefont {Shi}}]{Wang15}%
  \BibitemOpen
  \bibfield  {author} {\bibinfo {author} {\bibfnamefont {Z.}~\bibnamefont
  {Wang}}, \bibinfo {author} {\bibfnamefont {C.}~\bibnamefont {Tang}}, \bibinfo
  {author} {\bibfnamefont {R.}~\bibnamefont {Sachs}}, \bibinfo {author}
  {\bibfnamefont {Y.}~\bibnamefont {Barlas}}, \ and\ \bibinfo {author}
  {\bibfnamefont {J.}~\bibnamefont {Shi}},\ }\href {\doibase
  10.1103/PhysRevLett.114.016603} {\bibfield  {journal} {\bibinfo  {journal}
  {Phys. Rev. Lett.}\ }\textbf {\bibinfo {volume} {114}},\ \bibinfo {pages}
  {016603} (\bibinfo {year} {2015})}\BibitemShut {NoStop}%
\bibitem [{\citenamefont {Nomura}\ and\ \citenamefont
  {MacDonald}(2007)}]{Nomura2007}%
  \BibitemOpen
  \bibfield  {author} {\bibinfo {author} {\bibfnamefont {K.}~\bibnamefont
  {Nomura}}\ and\ \bibinfo {author} {\bibfnamefont {A.~H.}\ \bibnamefont
  {MacDonald}},\ }\href {http://link.aps.org/doi/10.1103/PhysRevLett.98.076602}
  {\bibfield  {journal} {\bibinfo  {journal} {Physical Review Letters}\
  }\textbf {\bibinfo {volume} {98}},\ \bibinfo {pages} {076602} (\bibinfo
  {year} {2007})}\BibitemShut {NoStop}%
\bibitem [{\citenamefont {Tzalenchuk}\ \emph {et~al.}(2010)\citenamefont
  {Tzalenchuk}, \citenamefont {Lara-Avila}, \citenamefont {Kalaboukhov},
  \citenamefont {Paolillo}, \citenamefont {Syv{\"a}j{\"a}rvi}, \citenamefont
  {Yakimova}, \citenamefont {Kazakova}, \citenamefont {Janssen}, \citenamefont
  {Fal'Ko},\ and\ \citenamefont {Kubatkin}}]{tzalenchuk2010towards}%
  \BibitemOpen
  \bibfield  {author} {\bibinfo {author} {\bibfnamefont {A.}~\bibnamefont
  {Tzalenchuk}}, \bibinfo {author} {\bibfnamefont {S.}~\bibnamefont
  {Lara-Avila}}, \bibinfo {author} {\bibfnamefont {A.}~\bibnamefont
  {Kalaboukhov}}, \bibinfo {author} {\bibfnamefont {S.}~\bibnamefont
  {Paolillo}}, \bibinfo {author} {\bibfnamefont {M.}~\bibnamefont
  {Syv{\"a}j{\"a}rvi}}, \bibinfo {author} {\bibfnamefont {R.}~\bibnamefont
  {Yakimova}}, \bibinfo {author} {\bibfnamefont {O.}~\bibnamefont {Kazakova}},
  \bibinfo {author} {\bibfnamefont {T.}~\bibnamefont {Janssen}}, \bibinfo
  {author} {\bibfnamefont {V.}~\bibnamefont {Fal'Ko}}, \ and\ \bibinfo {author}
  {\bibfnamefont {S.}~\bibnamefont {Kubatkin}},\ }\href@noop {} {\bibfield
  {journal} {\bibinfo  {journal} {Nature nanotechnology}\ }\textbf {\bibinfo
  {volume} {5}},\ \bibinfo {pages} {186} (\bibinfo {year} {2010})}\BibitemShut
  {NoStop}%
\bibitem [{\citenamefont {Jeckelmann}\ and\ \citenamefont
  {Jeanneret}(2001)}]{jeckelmann2001quantum}%
  \BibitemOpen
  \bibfield  {author} {\bibinfo {author} {\bibfnamefont {B.}~\bibnamefont
  {Jeckelmann}}\ and\ \bibinfo {author} {\bibfnamefont {B.}~\bibnamefont
  {Jeanneret}},\ }\href@noop {} {\bibfield  {journal} {\bibinfo  {journal}
  {Reports on Progress in Physics}\ }\textbf {\bibinfo {volume} {64}},\
  \bibinfo {pages} {1603} (\bibinfo {year} {2001})}\BibitemShut {NoStop}%
\end{thebibliography}%

\end{document}